\documentclass[english,aps,prc,floats,floatfix,nofootinbib]{revtex4}
\usepackage[T1]{fontenc}
\usepackage[latin9]{inputenc}
\usepackage{float}
\usepackage{graphicx}

\makeatletter
\providecommand{\tabularnewline}{\\}

\usepackage{color}

\@ifundefined{textcolor}{}
{
 \definecolor{BLACK}{gray}{0}
 \definecolor{WHITE}{gray}{1}
 \definecolor{RED}{rgb}{1,0,0}
 \definecolor{GREEN}{rgb}{0,1,0}
 \definecolor{BLUE}{rgb}{0,0,1}
 \definecolor{CYAN}{cmyk}{1,0,0,0}
 \definecolor{MAGENTA}{cmyk}{0,1,0,0}
 \definecolor{YELLOW}{cmyk}{0,0,1,0}
 \definecolor{DARKGREEN}{rgb}{0.,0.5,0.}
}

\usepackage{bm}
\makeatletter
\usepackage{graphics}
\usepackage{epsf}

\makeatother

\def\s#1{\setbox0=\hbox{$#1$}  
   \dimen0=\wd0     
   \setbox1=\hbox{/} \dimen1=\wd1  
   \ifdim\dimen0>\dimen1   
      \rlap{\hbox to \dimen0{\hfil/\hfil}} 
      #1     
   \else     
      \rlap{\hbox to \dimen1{\hfil$#1$\hfil}} 
      /      
   \fi}      %

\usepackage{babel}

\begin{document}

\title{Octet-baryon axial-vector charges and $SU(3)$-breaking
effects in the semileptonic hyperon decays}

\author{T. Ledwig$^{1}$}

\email{ledwig@ific.uv.es}

\author{J. Martin Camalich$^{2,3}$}

\email{jmartincamalich@ucsd.edu}

\author{L. S. Geng$^{4}$}

\email{lisheng.geng@buaa.edu.cn}

\author{M. J. Vicente Vacas$^{1}$}

\email{vicente@ific.uv.es}

\affiliation{$^{1}$Departamento de Física Teórica and IFIC, Centro Mixto, 
Institutos de Investigación de Paterna - Universidad de Valencia-CSIC, Spain\\
$^2$Dept. Physics, University of California, San Diego, 9500 Gilman Drive, 
La Jolla, CA 92093-0319, USA\\
$^3$PRISMA Cluster of Excellence Institut f\"ur Kernphysik, 
Johannes Gutenberg-Universit\"at Mainz, 55128 Mainz, Germany\\
$^4$School of Physics and Nuclear Energy Engineering and International 
Research Center for Particles and Nuclei in the Cosmos, Beihang University,  
Beijing 100191,  China}

\begin{abstract}
The octet-baryon axial-vector charges and the $g_{1}/f_{1}$ ratios measured 
in the semileptonic hyperon decays are studied up to $\mathcal{O}(p^{3})$ 
using the covariant baryon chiral perturbation theory with explicit decuplet
contributions. We clarify the role of different low-energy constants 
and find a good convergence for the chiral expansion of the axial-vector charges of the baryon octet, 
$g_1(0)$, with $\mathcal{O}(p^3)$ corrections typically 
around $20\%$ of the leading ones. This is a consequence of 
strong cancellations between different next-to-leading order terms. We 
show that considering only non-analytic terms is not 
enough and that analytic terms appearing at the same chiral order
play an important role in this description. The same 
effects still hold for the chiral extrapolation of the axial-vector 
charges and result in a rather mild quark-mass dependence.
As a result, we report a determination of the leading order
chiral couplings, $D=0.623(61)(17)$ and $F=0.441(47)(2)$, as obtained 
from a completely consistent chiral analysis up to $\mathcal{O}(p^{3})$. 
Furthermore, we note that the appearance of an unknown low-energy constant
precludes the extraction of the proton octet-charge from 
semileptonic decay data alone, which is relevant for an analysis 
of the composition of the proton spin.
\end{abstract}



\maketitle

\section{Introduction}
The non-perturbative regime of QCD is dominated by the spontaneous breaking of 
the chiral symmetry. Based on that, an effective field theory of QCD 
at low-energies is constructed using the pseudo-scalar mesons
and baryons as basic degrees of freedom.
This theory is called baryon chiral perturbation theory (B$\chi$PT) 
\cite{Weinberg:1978kz,Gasser:1983yg, Gasser:1984gg, Gasser:1987rb}, 
and it parametrizes the axial-vector 
(AV) structure of the octet baryons and the meson-baryon interaction at 
leading order (LO) by the only two low-energy constants (LECs), $D$ and $F$. 
These are essential parameters in this model-independent 
approach and they are one of the main topics of this work.

A reliable experimental source to determine $D$ and $F$ are the ratios 
of the axial-vector and vector couplings, $g_{1}/f_{1}$, as measured in the
semileptonic hyperon decays (SHD)\footnote{In the $SU\left(2\right)$ version 
of B$\chi$PT, only the combination $D+F$ is accessible, which is at leading 
order equal to the AV charge of the nucleon $g_{A}=1.2701\left(25\right)
\times g_V$ as measured in neutron $\beta$-decay \cite{Beringer:1900zz}.}. 

Already several decades ago Cabibbo proposed a $SU(3)$ symmetric model
\cite{Cabibbo:1963yz} for the weak hadronic currents.  A fit of this model to 
the current data is very successful, yielding $D\approx0.804$ 
and $F\approx0.463$, and implying that $SU(3)$ 
symmetry breaking effects in SHD are small \cite{Cabibbo:2003cu}. 
Supporting this interpretation, the experimental measurements of 
$g_{1}/f_{1}$ in the $n\to pe\overline{\nu}$ and 
$\Xi^{0}\to\Sigma^{+}e\overline{\nu}$ decays, 
which are predicted by this model to be exactly equal, differ only
by a $\sim5\%$ \cite{AlaviHarati:2001xk,Batley:2006fc,Beringer:1900zz}. 

From a modern perspective the success of the Cabibbo model is intriguing 
given that the $SU(3)$-flavor symmetry is explicitly broken by 
$m_s\gg m_u\sim m_d$. For instance, in B$\chi$PT this model corresponds 
to the LO approximation while nearly all higher order corrections break 
the $SU(3)$ symmetry. As a consequence, the next-to-leading 
order (NLO) contributions
must arrange themselves in such a way that the net breaking effects remain 
small. Additionally, the total NLO effect has also to be 
small compared to the LO one for the chiral 
expansion to make sense. 

These issues were discussed in the foundational papers of the heavy-baryon
(HB)$\chi$PT approach \cite{Jenkins:1990jv,Jenkins:1991es}, where it was 
found that the NLO chiral corrections 
to the AV charges can be large and problematic. However, a 
cancellation mechanism between loops with intermediate octet- and 
decuplet-baryons was revealed and showed to produce a reasonable 
description of the data and convergence of the chiral series. This 
was later found to be a consequence of the $SU(6)$ spin-flavor symmetry 
that emerges in the large $N_c$ limit of the baryonic sector of QCD. Thus, 
much of the subsequent work on the axial structure has focused on the combination
of HB$\chi$PT and Large $N_c$ to ensure the octet-decuplet cancellations at each
level of the perturbative expansion 
\cite{Dashen:1993as,Dashen:1993jt,Dashen:1994qi,FloresMendieta:2000mz,
FloresMendieta:2012dn,CalleCordon:2012xz}.

Nevertheless, from the point of view of the chiral expansion all the 
early and later works in HB$\chi$PT were not entirely systematic as they 
focused on the loop corrections but neglected the effects of various 
local operators appearing at NLO. In fact, there is a total of 
six new LECs that contribute to the AV charge in SHDs at this 
order. Four of them break $SU(3)$ whereas the other two have the same 
structure as $D$ and $F$ but come multiplied by a singlet combination 
of quark masses. As a result, one can absorb the latter into $D$ and $F$, 
and fit the resulting six LECs to the six available measurements 
of $g_{1}/f_{1}$. Such a study has been carried out in the infrared 
(IR) scheme of covariant B$\chi$PT \cite{Becher:1999he,Zhu:2000zf}, 
and it was shown that the recoil corrections
 included in the relativistic calculation of the loops in this approach 
could be as large as the LO contributions. The main conclusion of this work 
was that the chiral expansion of AV charges is not convergent \cite{Zhu:2000zf} .

These findings and, in general, the analysis of the AV 
couplings in B$\chi$PT need to be revisited. In the first place, the IRB$\chi$PT
employed in the latter work is known to introduce spurious cuts that can have 
important effects in phenomenology
~\cite{Geng:2008mf,Ledwig:2010nm,Ledwig:2011cx,Alarcon:2012kn}. 
Secondly, the decuplet contributions were 
neglected despite the fact the typical octet-decuplet mass splitting, 
$(M_{\Delta}-M_{N})/\Lambda_{SB}\approx0.3$, is smaller than the perturbation 
$M_{K}/\Lambda_{SM}\approx0.5$ and their effects provide the important source of 
cancellations at NLO induced by the symmetries of QCD at Large $N_c$. 
Finally, the absorption of the two $\mathcal{O}(p^3)$ singlet LECs into $D$ and $F$
precludes a definite discussion on the chiral convergence as these 
contributions appear at different orders.

In this work we analyze the AV charges of the baryons 
in a completely consistent fashion within B$\chi$PT 
and put the description of the experimental $g_1/f_1$ ratios on a 
systematic ground. We employ the extended-on-mass-shell
renormalization scheme (EOMS) \cite{Gegelia:1999gf,Fuchs:2003qc}, 
which is a relativistic solution to the power counting problem found in 
\cite{Gasser:1987rb}
that leaves the analytic structure of the relativistic loops intact. To include
explicit decuplet contributions and to ensure the decoupling of the
spurious spin-1/2 components of the spin-3/2 Rarita-schwinger fields,
we use the consistent couplings of 
\cite{Pascalutsa:1998pw,Pascalutsa:1999zz,Pascalutsa:2000kd,Pascalutsa:2006up,Geng:2009hh}.
In contrast to the IRB$\chi$PT study \cite{Zhu:2000zf}, we do not 
absorb $\mathcal{O}(p^3)$ LECs in $D$ and $F$. In order to disentangle 
the two singlet LECs we use the recent $N_f=2+1$ lattice 
QCD (lQCD) calculations \cite{Lin:2007ap,Gockeler:2011ze} of the isovector 
AV constants $g_{A}^{3}$ of the proton, $\Sigma^{+}$ and $\Xi^{0}$. 
These are additional data points which we include in our fits along 
with the experimental SHD data.

We report that B$\chi$PT at $\mathcal{O}(p^3)$ successfully
describes the AV charges of the baryon octet. The NLO 
corrections are typically about $\sim 20\%$ of the LO ones, 
which is consistent with the expectations for a convergent expansion. 
We extract $D$ and $F$ at this order and we discuss further implications 
of our study for the structure of the spin of the proton.   

The work is organized as follows. Section two defines the AV form
factors and gives the measured transitions used as fit input. The
third section introduces the covariant $SU(3)$ B$\chi$PT
with explicit decuplet degrees of freedom and the EOMS renormalization
scheme. In the fourth section we present and discuss the results of
our SHD study. The fifth section summarizes our work and relevant
technical expressions are given in the appendices.

\section{semileptonic hyperon decays and axial-vector form factors
  \label{sec:semileptonic-hyperon-decays}}

The AV structure of the baryon octet can be accessed via 
the $\beta$-decays of hyperons, $B\to B^{\prime}e\overline{\nu}_{e}$. 
We parametrize the decay amplitude as~\cite{Cabibbo:2003cu}
\begin{equation}
\mathcal{M}=\frac{G}{\sqrt{2}}\overline{u}^{\prime}\left(p^{\prime}\right)\left[\mathcal{O}_{V\left(B^{\prime}B\right)}^{\alpha}\left(p^{\prime},p\right)+\mathcal{O}_{A\left(B^{\prime}B\right)}^{\alpha}\left(p^{\prime},p\right)\right]u\left(p\right)\,\,\overline{u}_{e}\left(p_{e}\right)\left[\gamma_{\alpha}+\gamma_{\alpha}\gamma^{5}\right]v_{\nu}\left(p_{\nu}\right)\,\,\,\,,
\end{equation}
with $u\left(p\right)$, $\overline{u}^{\prime}\left(p^{\prime}\right)$
the spin-1/2 spinors for the baryons $B$ and $B^{\prime}$ with momenta
$p$, $p^{\prime}$ and $\overline{u}_{e}\left(p_{e}\right)$, 
$v_{\nu}\left(p_{\nu}\right)$
as the electron and anti-neutrino spinors with momenta $p_{e}$, 
$p_{\nu}$. The coupling $G$ is defined by $G=G_{F}V_{ud}$ for the 
strangeness-conserving and $G=G_{F}V_{us}$ for the strangeness-changing 
processes with $|\Delta S|=1$, where $G_{F}$ and $V_{ud\left(us\right)}$ are the
Fermi coupling constant and the respective CKM matrix elements. Using 
parity-invariance arguments, both the vector and AV operators 
$\mathcal{O}_{V\left(B^{\prime}B\right)}^{\alpha}\left(p^{\prime},p\right)$
and $\mathcal{O}_{A\left(B^{\prime}B\right)}^{\alpha}\left(p^{\prime},p\right)$
contain three independent Lorentz-structures 
\begin{eqnarray}
\mathcal{O}_{V\left(B^{\prime}B\right)}^{\alpha}\left(p^{\prime},p\right) & = & f_{1}^{B^{\prime}B}\left(q^{2}\right)\gamma^{\alpha}-\frac{i}{M_{B}}\sigma^{\alpha\beta}q_{\beta}f_{2}^{B^{\prime}B}\left(q^{2}\right)+\frac{1}{M_{B}}q^{\alpha}f_{3}^{B^{\prime}B}\left(q^{2}\right)\,\,\,\,,\\
\mathcal{O}_{A\left(B^{\prime}B\right)}^{\alpha}\left(p^{\prime},p\right) & = & g_{1}^{B^{\prime}B}\left(q^{2}\right)\gamma^{\alpha}\gamma^{5}-\frac{i}{M_{B}}\sigma^{\alpha\beta}q_{\beta}\gamma^{5}g_{2}^{B^{\prime}B}\left(q^{2}\right)+\frac{1}{M_{B}}q^{\alpha}\gamma^{5}g_{3}^{B^{\prime}B}\left(q^{2}\right)\,\,\,\,,
\end{eqnarray}
with $\sigma^{\alpha\beta}=i\left[\gamma^{\alpha},\gamma^{\beta}\right]/2$
and $q^{\alpha}=\left(p^{\prime}-p\right)^{\alpha}$ and $f_{i}$,
$g_{i}$ as the vector and AV form factors, normalized by
the mass $M_{B}$ of the baryon $B$. These functions contain information about
the internal structure of the baryons as probed by AV sources. 

The quantities we study in this work are the AV charges 
$g_{1}^{B^{\prime}B}\left(q^{2}=0\right)\equiv g_{1}^{B^{\prime}B}$.
They are part of the ratios $g_{1}\left(0\right)/f_{1}\left(0\right)\equiv g_{1}/f_{1}$
which are measured through the SHD. 
The $SU(3)$ breaking corrections to the vector charges are of a 
few percent \cite{Ademollo:1964sr,Geng:2009ik,Sasaki:2012ne,Geng:2014efa} 
and can be safely neglected at the NLO accuracy in the chiral expansion of $g_{1}/f_{1}$. 
Thus, we use the $SU(3)$ symmetric values for $f_{1}\left(0\right)\equiv f_{1}$ to 
extract experimental values for $g_1$.

In Tab. \ref{tab:Semileptonic-hyperon-data} we list the only six
measured SHD processes which are not related by isospin symmetry,
as e.g. $f_{1}^{\Xi^{0}\Sigma^{+}}=\sqrt{2}f_{1}^{\Xi^{-}\Sigma^{0}}$ and
$g_{1}^{\Xi^{0}\Sigma^{+}}=\sqrt{2}g_{1}^{\Xi^{-}\Sigma^{0}}$.
The data is taken from \cite{Beringer:1900zz}, where a 
different notation for the $\beta$-decay is used, 
which results in a different sign of the $g_{1}$ definition. For
the sign of the mode $\Xi^{0}\to\Sigma^{+}$ we also refer to Refs.
\cite{AlaviHarati:2001xk,Batley:2006fc}. Furthermore, we list the $SU(3)$
symmetric values for the $f_{1}$ and $g_{1}$ results of the Cabibbo
model \cite{Cabibbo:2003cu}, which are equivalent to the B$\chi$PT at 
LO. Finally, for the $\Sigma^{-}\to\Lambda e^{-}\overline{\nu}$ channel 
$f_1=0$ up to $\mathcal{O}\left( (m_d-m_u)^2 \right)$ and $g_1$ can be determined directly 
from the total decay rate \cite{Beringer:1900zz}.

\begin{table}
  \caption{\label{tab:Semileptonic-hyperon-data}Upper table: Semileptonic hyperon
    data for the decays $B\to B^{\prime}e^{-}\overline{\nu}_{e}$. The
    values are taken from \cite{Beringer:1900zz} where the experimental result for
    $\Sigma^{-}\to\Lambda$ is obtained as described in the text. The last 
    two rows correspond to the $SU(3)$ symmetric values of the Cabibbo model. 
    Lower table: The lQCD data from \cite{Lin:2007ap,Gockeler:2011ze} entering our fits for the
    AV charges $g_{A,3}^{X}$ for the proton (P), $\Sigma^{+}$ and $\Xi^{0}$. Note that the normalization
    of $\Sigma^{+}$ in \cite{Lin:2007ap} is half the one
    used here. For $M_{\eta}$ we use the Gell-Mann-Okubo mass relation.}
  
  \begin{center}
    \begin{tabular}{ccccccc}
      \hline 
      \hline 
      & $n\to p$ & $\Lambda\to p$ & $\Sigma^{-}\to n$ & $\Sigma^{-}\to\Lambda$
      \footnote{Since $f_{1}=0$, we list $\sqrt{3/2}\, g_{1}$ instead of $g_{1}/f_{1}$.}
      & $\Xi^{0}\to\Sigma^{+}$ & $\Xi^{-}\to\Lambda$\tabularnewline
      \hline 
      $g_{1}/f_{1}$  & $1.270(3)$ & $0.718(15)$ & $-0.340(17)$ & $0.698(33)$ & $1.210(50)$ & $0.250(50)$
      \tabularnewline
      $f_{1}^{SU(3)}$ & $1$ & $-\sqrt{\frac{3}{2}}$ & $-1$ & $0$ & $1$ & $\sqrt{\frac{3}{2}}$
      \tabularnewline
      $g_{1}^{SU(3)}$ & $D+F$ & $\frac{-1}{\sqrt{6}}(D+3F)$ & $D-F$ & $\sqrt{\frac{2}{3}}D$ & $D+F$ & $\frac{-1}{\sqrt{6}}(D-3F)$\tabularnewline
      \hline 
      \hline 
    \end{tabular}
  \end{center}
  
  \begin{tabular}{ccccccc}
    \hline 
    \hline 
    $P$ & $\Sigma^{+}$ & $\Xi^{0}$ & $\frac{\Sigma^{+}}{P}$ & $\frac{\Xi^{0}}{P}$ & $M_{\pi}$ {[}MeV{]} & $M_{K}$ {[}MeV{]}\tabularnewline
    $1.22$ & $2\times0.418$ & $-0.262$ &  &  & $354$ & $604$\tabularnewline
    &  &  & $0.76$ & $-0.23$ & $350$ & $485$\tabularnewline
    &  &  & $0.76$ & $-0.22$ & $377$ & $473$\tabularnewline
    &  &  & $0.78$ & $-0.22$ & $414$ & $459$\tabularnewline
    &  &  & $0.77$ & $-0.23$ & $443$ & $443$\tabularnewline
    &  &  & $0.78$ & $-0.23$ & $481$ & $420$\tabularnewline
    \hline 
    \hline 
  \end{tabular}

\end{table}

In addition to the experimental data, we use also lQCD results from $N_f=2+1$ 
ensembles for the isovector AV charges of the proton, $\Sigma$ and $\Xi$. 
Introducing these results at different
non-physical quark masses allows for separating the LO parameters $D$ and $F$
from other $p^3$ LECs. In particular, we include the lowest $M_{\pi}$ 
data points from the Hadron-Spectrum collaboration \cite{Lin:2007ap} as well as 
the whole set of the AV ratios for $\Sigma^{+}/P$ and $\Xi^{0}/P$ 
from the QCDSF-UKQCD collaboration \cite{Gockeler:2011ze}.
The latter study is done along the $SU(3)$ singlet
line where the quantity $2M_{k}^{2}+M_{\pi}^{2}$
is kept constant with the pion and kaon masses each chosen to be 
smaller than the physical kaon mass. We individually list all these data points in 
Tab. \ref{tab:Semileptonic-hyperon-data}.
However, we have to note that the AV coupling of 
the proton is known to suffer from not fully-understood lattice artifacts
~\cite{Green:2012ud,Capitani:2012gj}. 
Therefore, we increase the lQCD uncertainties to be a $\sim10\%$ relative to the central 
values, which is roughly a factor 5 larger than the errors usually quoted. 
We assume this accounts for lattice systematic effects such as 
excited-state contamination, finite-volume or discretization corrections 
which will not be addressed in this work.

For the isovector AV form factors we use the following parametrization:  
\begin{equation}
  \langle B\left(p^{\prime}\right)|\overline{q}\gamma^{\alpha}\gamma^{5}\lambda^{3}q|B\left(p\right)\rangle=\overline{u}_{B}^{\prime}\left(p^{\prime}\right)\left[G_{A,BB}^{3}\left(q^{2}\right)\gamma^{\alpha}+\frac{1}{2M_{B}}G_{P,BB}^{3}\left(q^{2}\right)q^{\alpha}\right]\gamma^{5}u_{B}\left(p\right)\,\,\,,
\end{equation}
with $\lambda^{3}$ as a Gell-Mann matrix and 
$G_{A,BB}^{3}\left(q^{2}=0\right)\equiv g_{A,3}^{BB}$
as the isovector AV constant and $G_{P,BB}^{3}\left(p^{2}\right)$
the induced pseudo-scalar form factor. It is worth recalling that these form 
factors are related by isospin symmetry to those appearing in the $\beta$-decays
 $n\rightarrow p$, $\Sigma^-\rightarrow\Sigma^0$, $\Sigma^0\rightarrow\Sigma^+$ 
and $\Xi^-\rightarrow\Xi^0$.    

\section{Baryon chiral perturbation theory \label{BChPT}}

Chiral perturbation theory ($\chi$PT) allows for model-independent and systematic 
studies of hadronic phenomena in the low-energy regime of QCD. It consists of a 
perturbative expansion in $p/\Lambda_{SB}$ where $\Lambda_{SB}=4\pi f_{\pi}\approx1$ 
GeV is the scale of the spontaneous chiral symmetry breaking and $p$ is either the 
typical energy involved in the process or the quark-masses which break the chiral 
symmetry explicitly~
\cite{Weinberg:1978kz,Gasser:1983yg, Gasser:1984gg}. 
Only chiral-symmetry arguments are used to construct the 
effective Lagrangian. The free LECs appearing with the different operators 
must be determined using nonperturbative calculations in QCD (e.g. lQCD) 
or experimental data. 

The extension of $\chi$PT to the baryon sector implies some difficulties. 
One is that the baryon mass introduces a new hard scale which leads 
to the breakdown of the naive power counting~\cite{Gasser:1987rb}.
This can be solved by integrating out these hard modes from the outset, like in 
HB$\chi$PT, although the recoil corrections to the loop functions, incorporated 
order-by-order in the HB expansion, can be large, especially in $SU(3)$
~\cite{Geng:2008mf,Geng:2009hh,MartinCamalich:2010fp,Geng:2013xn}. Alternatively, one can use
a manifestly covariant formulation exploiting the fact that all the power-counting 
breaking terms are analytic~\cite{Becher:1999he}. 
Therefore, they have the same structure as the local 
operators of the most general chiral Lagrangian and can be cancelled by a suitable 
renormalization prescription. Two schemes stand out among the manifestly covariant 
formalisms, the IRChPT~\cite{Becher:1999he} and the EOMS ChPT
~\cite{Gegelia:1999gf,Fuchs:2003qc}. The IRChPT~\cite{Becher:1999he} uses a 
regularization procedure which has been shown to alter the analytic structure of 
the loops and to spoil the description of some observables
~\cite{Geng:2008mf,Ledwig:2010nm,Ledwig:2011cx,Alarcon:2012kn}. On the other hand, 
the EOMS scheme is a minimal-subtraction scheme in which the finite parts of the 
available bare LECs cancel the power-counting-breaking terms
~\cite{Gegelia:1999gf,Fuchs:2003qc}. 
This procedure has the advantage that it incorporates the recoil corrections of 
the loops graphs to all orders in consistency with analyticity.  A second difficulty 
in B$\chi$PT is related to the closeness in mass of the decuplet resonances. Indeed, 
the octet-decuplet mass splitting $\Delta$ is about 300 MeV, which is smaller than the 
maximal scale of perturbations $M_K\sim495$ MeV, and the decuplet 
baryons should be introduced as dynamical degrees of freedom in the framework.    

\begin{figure}
  \caption{\label{fig: g1 diagrams} Feynman diagrams contributing up to 
    $\mathcal{O}\left(p^{3}\right)$ to the $g_{1}^{B^{\prime}B}$ AV 
    form factor. Single solid lines denote octet baryons and double lines 
    decuplet baryons. The dashed lines correspond to mesons and the wiggly 
    line to the external AV field. A number inside the vertex 
    denote its chiral order.}

  \begin{center}
  \includegraphics[scale=0.33]{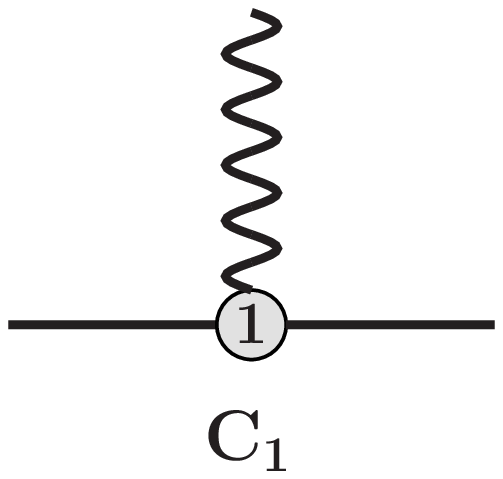}~~\includegraphics[scale=0.33]{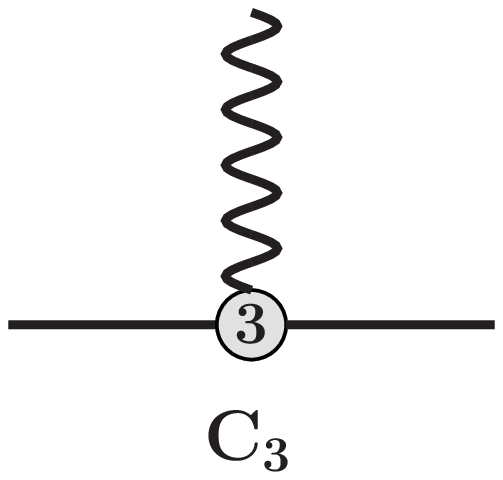}
  ~~\includegraphics[scale=0.33]{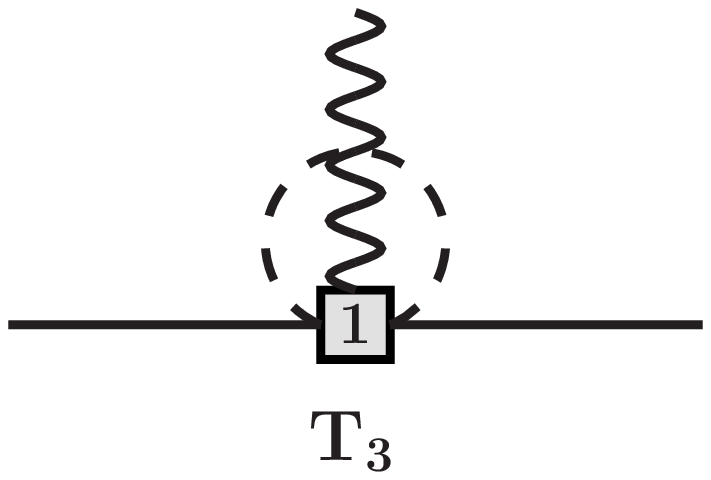}~~\includegraphics[scale=0.33]{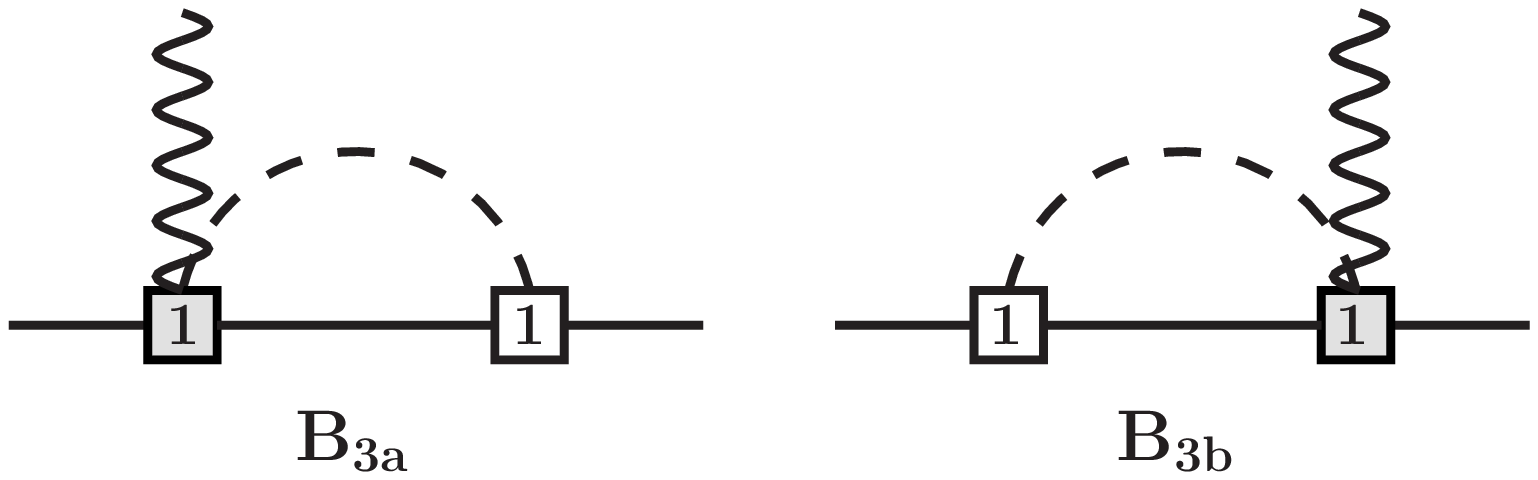}
  ~~\includegraphics[scale=0.33]{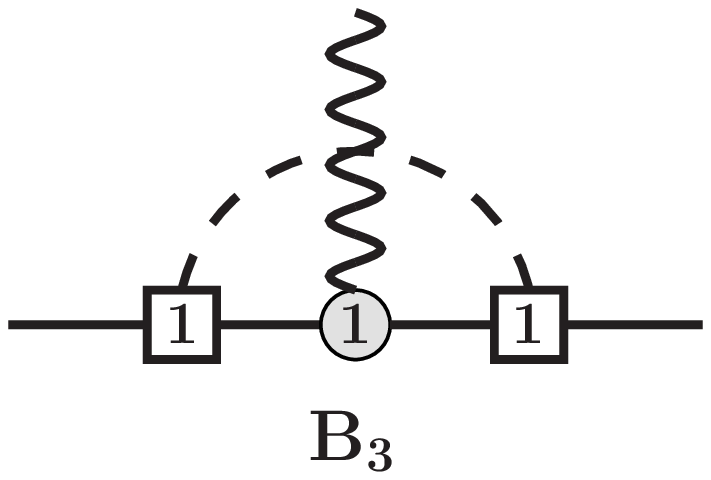}\\
  \includegraphics[scale=0.33]{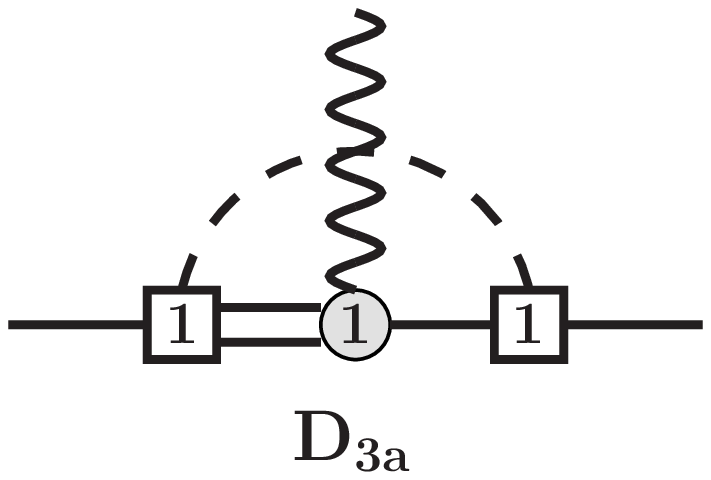}~~\includegraphics[scale=0.33]{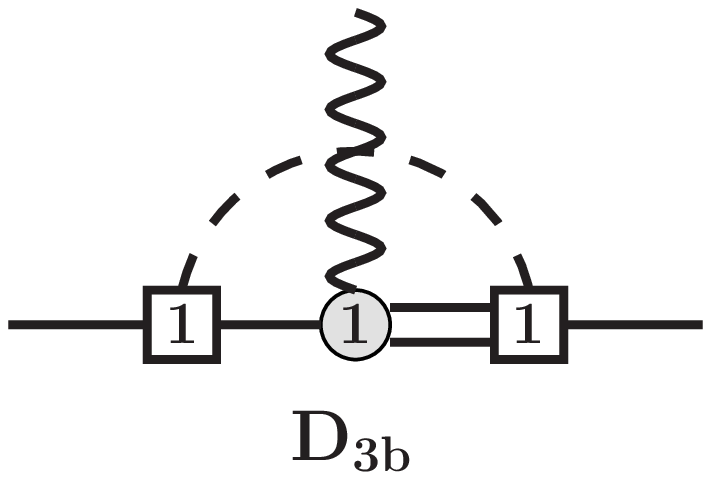}
  ~~\includegraphics[scale=0.33]{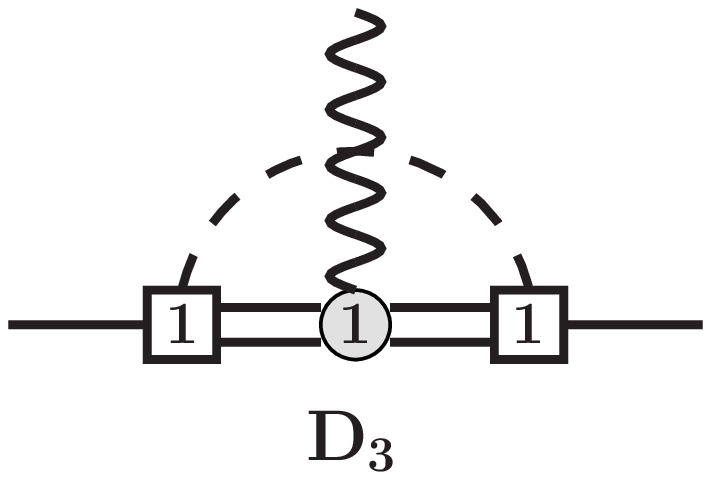}
  \end{center}

\end{figure}

In this work we employ the covariant $SU(3)$ B$\chi$PT up to order
$p^{3}$ with inclusion of explicit decuplet degrees of freedom and
the EOMS renormalization scheme \cite{Gegelia:1999gf,Fuchs:2003qc}. 
The field content of the theory
 are the octet baryons, $B\left(x\right)$,
and decuplet baryons, $T\left(x\right)$, interacting with the pseudo-scalar
octet $\phi\left(x\right)$ and an external AV field 
$a_{\mu}\left(x\right)$. We use an equivalent of the small-scale-expansion scheme 
\cite{Hemmert:1996xg} to count $p\sim M_{\phi}\sim\Delta\sim\epsilon$,
denoting all small scales commonly by $\epsilon$.
Accordingly, the chiral order $n$ of a Feynman graph is given by 
\begin{equation}
  n=4L-2N_{\phi}-N_{B}-N_{D}+\sum_{k}kV_{k}\,\,\,\,,\label{eq: Power Counting}
\end{equation}
for a graph with $L$ loops, $N_{\phi}$ internal mesons, $N_{B}$
internal octet baryons, $N_{D}$ internal decuplet baryons and $V_{k}$
vertices from a $\mathcal{L}^{\left(k\right)}$ Lagrangian. Using
Eq. (\ref{eq: Power Counting}) together with the Lagrangian and the
renormalization scheme specified below, we list in Fig. \ref{fig: g1 diagrams}
all Feynman graphs that contribute to the AV charges up
to order $p^{3}$. 

For our study, we need the following four terms from the B$\chi$PT Lagrangian
\begin{equation}
  \mathcal{L}=\mathcal{L}_{B}^{\left(1\right)}+\mathcal{L}_{B}^{\left(3\right)}+\mathcal{L}_{D}^{\left(1\right)}+\mathcal{L}_{BD}^{\left(1\right)}\,\,\,\,,\label{eq:Lfull}
\end{equation}
where the last two contributions contain the decuplet fields.
The number in brackets denotes the chiral order of each part. The
first term is the standard leading-order baryon-octet Lagrangian and
the second term the $p^{3}$-order part constructed in 
\cite{Oller:2006yh,Frink:2006hx,Oller:2007qd}.
Their explicit expressions are 
\begin{eqnarray}
  \mathcal{L}_{B}^{\left(1\right)} & = & \langle\overline{B}\left(i\s D-M_{B0}\right)B\rangle+\frac{D}{2}\langle\overline{B}\gamma^{\mu}\gamma_{5}\left\{ u_{\mu},B\right\} \rangle+\frac{F}{2}\langle\overline{B}\gamma^{\mu}\gamma_{5}\left[u_{\mu},B\right]\rangle\,\,\,,\label{eq:L1B}\\
  \mathcal{L}_{B}^{\left(3\right)} & = & +h_{38}\langle\overline{B}u^{\mu}\gamma_{\mu}\gamma_{5}B\chi_{+}\rangle+h_{39}\langle\overline{B}\chi_{+}\gamma_{\mu}\gamma_{5}Bu^{\mu}\rangle+h_{40}\langle\overline{B}u^{\mu}\gamma_{\mu}\gamma_{5}B\rangle\langle\chi_{+}\rangle\nonumber+h_{41}\langle\overline{B}\gamma_{\mu}\gamma_{5}Bu^{\mu}\rangle\langle\chi_{+}\rangle \\
  &  & +h_{42}\langle\overline{B}\gamma_{\mu}\gamma_{5}B\rangle\langle u^{\mu}\chi_{+}\rangle+h_{43}\langle\overline{B}\gamma_{\mu}\gamma_{5}B\left\{ u^{\mu},\chi_{+}\right\} \rangle+h_{44}\langle\overline{B}\left\{ u^{\mu},\chi_{+}\right\} \gamma_{\mu}\gamma_{5}B\rangle\nonumber+...\,\,\,\,,
\end{eqnarray}
where $\langle...\rangle$ denotes the flavor trace and all 
further notations are explained in App. \ref{App:Notation}.
All the LECs in the chiral Lagrangians are formally
defined in the chiral limit where, for instance, $M_{B0}$ 
represents the corresponding baryon mass.
At LO, the complete meson-baryon and AV
baryon interactions are parametrized by only the two LECs $D$ and $F$. 
The $\mathcal{L}_{B}^{\left(2\right)}$ does not contain operators that contribute 
to the AV couplings of the baryons, while several appear at 
$\mathcal{O}(p^3)$ that are parameterized by the $h_{i}$ LECs.
Note that here we choose the $h_{i}$ with the opposite sign as in \cite{Oller:2006yh,Frink:2006hx,Oller:2007qd}
 and that only the structures $h_{38,39}$ and $h_{43,44}$ contain explicit
$SU(3)$ symmetry breaking terms while the
structures $h_{40,41}$ include $SU(3)$-singlets. 
Finally, the LEC $h_{42}$ does not contribute to SHDs or the isovector couplings
(in the isospin limit), although it contributes to the singlet and octet 
charges of the baryons. We will discuss in Sec. \ref{sec:proton spin} 
the important consequences of this on the interpretation of the proton's spin.

For the decuplet Lagrangians we use:

\begin{eqnarray}
  \mathcal{L}_{D}^{\left(1\right)} & = & \overline{T}_{\mu}^{abc}\left[\gamma^{\mu\nu\alpha}i\partial_{\alpha}-M_{D0}\gamma^{\mu\nu}\right]T_{\nu}^{abc}-\frac{\mathcal{H}}{2 M_{D0}^{2}}\left(\partial_{\sigma}\overline{T}_{\tau}^{abi}\right)\gamma^{\alpha\sigma\tau}u_{\mu}^{ij}\gamma^{\mu}\gamma^{5}\gamma_{\alpha\kappa\lambda}\left(\partial^{\kappa}T^{abj\,\lambda}\right)\,\,\,\,,\label{eq:LDaD}\\
  \mathcal{L}_{DB}^{\left(1\right)} & = & \frac{i\mathcal{C}}{M_{D0}}\left[\left(\partial_{\mu}\overline{T}_{\nu}^{ijk}\right)\gamma^{\mu\nu\lambda} u_\lambda^{jl} B^{km}+\overline{B}^{mk}\gamma^{\mu\nu\lambda} u_\lambda^{lj} \left(\partial_{\mu}T_{\nu}^{ijk}\right)\right]\varepsilon^{ilm}\,\,\,\,,
\end{eqnarray}
where $X^{ab}$ denotes the matrix element in the $a$-th row and
$b$-th column. Each entry of the totally symmetric tensor $T^{abc}$
is a spin-3/2 Rarita-Schwinger spinor representing a decuplet baryon.
In App. \ref{App:Notation} we define explicitly all relevant quantities.
The $\mathcal{C}$ and $\mathcal{H}$ are the AV 
octet-decuplet and decuplet couplings, respectively, and $M_{D0}$ is the
chiral-limit decuplet baryon mass. In the case of $\mathcal{C}$, our definition
differs by a factor of $2$ as compared to the large $N_c$ work \cite{Dashen:1994qi}.

The above decuplet Lagrangians implement the \emph{consistent} couplings
of \cite{Pascalutsa:1998pw,Pascalutsa:1999zz,Pascalutsa:2000kd}. 
They are consistent in the sense that the
invariance of the free theory under a decuplet field redefinition
of $\Psi^{\mu}\to\Psi^{\mu}+\partial^{\mu}\epsilon\left(x\right)$,
with $\epsilon\left(x\right)$ a spinor field, carries over to the
interacting theory. This ensures the decoupling of the spurious spin-1/2
components of the Rarita-Schwinger spinor. In this way we also obtained the 
last term in Eq. (\ref{eq:LDaD}), i.e. by substituting
$\Psi^{\mu}\to\left(i/M_{D0}\right)\gamma^{\mu\alpha\nu}\left(\partial_{\alpha}\Psi_{\nu}\right)$
\cite{Pascalutsa:2006up} in the non-consistent Lagrangian 

\begin{equation}
  \mathcal{L}_{nc}^{\left(1\right)}=\frac{\mathcal{H}}{2}\overline{T}_{\alpha}^{abi} u_{\mu}^{ij} \gamma^{\mu}\gamma^{5}T^{abj\,\alpha}\,\,\,.
\end{equation}

With the above Lagrangians, we can now write down all terms that contribute
up to order $p^{3}$ to the AV charges $g_{1}^{B^{\prime}B}$
and $g_{A,3}^{BB}$. The full unrenormalized result in dimensional
regularization is 
\begin{eqnarray}
  g_{X}^{B^{\prime}B} & = & \sqrt{Z_{B^{\prime}}}\sqrt{Z_{B}}C_{1}^{B^{\prime}B}+C_{3}^{B^{\prime}B}+T_{3}^{B^{\prime}B}+B_{3}^{B^{\prime}B}+B_{3ab}^{B^{\prime}B}+D_{3}^{B^{\prime}B}+D_{3ab}^{B^{\prime}B}+\mathcal{O}\left(p^{4}\right)\,\,\,\,,\label{eq:g1BChPT}
\end{eqnarray}
where the notation matches the one of Fig. \ref{fig: g1 diagrams}
and we list all contributions explicitly in App. 
\ref{App:Axial-vector-form-factors}. The factors $Z_{B}$ are the 
wavefunction-renormalization constants which, at this order, only contribute 
through the LO terms. Furthermore, we apply the EOMS renormalization scheme 
\cite{Gegelia:1999gf,Fuchs:2003qc} at a scale $\Lambda=\overline{M}_{B}$.

We use Eq. (\ref{eq:g1BChPT}) to fit in Sec.~\ref{sec:Results} 
the data of Tab.~\ref{tab:Semileptonic-hyperon-data}. Some 
of the LECs appearing in the loop functions are already well 
constrained by other observables than the AV charges and we will 
use this additional information. Explicitly, these are the meson decay constant $f_{0}$, 
the baryon masses $M_{B0}$ and $M_{D0}$, and the couplings of the decuplet 
$\mathcal{C}$ and $\mathcal{H}$, all in the chiral limit. 
The former three can be determined using the extrapolation of
lQCD data, namely, $f_0 \simeq 87$ MeV~\cite{Aoki:2013ldr}, 
$M_{B0}\simeq880$~MeV~\cite{Ren:2012aj} and 
$M_{D0}\simeq1152$~MeV~\cite{Ren:2013dzt}. The decuplet couplings in the chiral
limit are not well known and we use the Large $N_c$ relations
$\mathcal{C}=-D$ and $\mathcal{H}=3D-9F$ 
\cite{Dashen:1994qi}, which are valid up to $1/N_c^2$ 
corrections. 

However, one can also use an alternative set for these parameters based on 
their experimental values which are better known. 
In this case, $f_{0}=\overline{f}$, 
$M_{0}=\overline{M}_{B}$ and $M_{D0}=\overline{M}_{D}$,
where $\overline{f}$ is the average of physical pion, kaon 
and $\eta$-decay constants and $\overline{M}_{B\left(D\right)}$ 
the average of the physical baryon masses in the respective 
multiplet. The octet-decuplet coupling is determined from the 
(strong) decuplet decays to $\mathcal{C}=-0.85(15)$. The 
experimental decuplet coupling $\mathcal{H}$ is not known and we 
use again the large $N_c$ relation. 

In Tab.~\ref{tab:Fit-parameter} we list the input parameters 
used in each case. Note that both choices are equivalent 
as one can rewrite one into the other at the expense of 
higher order contributions. We will perform our analysis 
with these two sets of values in order to assess systematic 
uncertainties.

As a final remark concerning the determination of $D$ and $F$ at 
$\mathcal{O}(p^3)$, we note that the LECs $h_{40,41}$ have the same structure 
as the LO couplings but come multiplied by a singlet of quark masses. We 
introduce lQCD results on the AV couplings in our statistical 
analysis precisely to disentangle these LECs from the $D$ and $F$.

\begin{figure}
  \caption{\label{fig: self-energies}Diagrams contributing to the wave-function
    renormalization. The notation is the same as in Fig. \ref{fig: g1 diagrams}.}
  \begin{center}
    \includegraphics[scale=0.33]{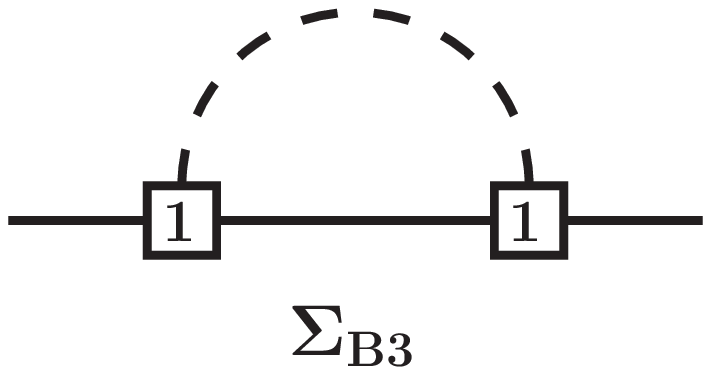}~~~~\includegraphics[scale=0.33]{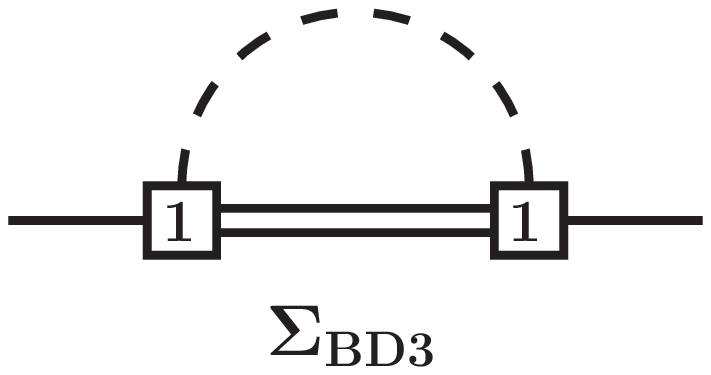}
  \end{center}
\end{figure}

\begin{table}
  \caption{\label{tab:Fit-parameter} Values of low-energy constants fixed in our fits. We list the
    meson decay constant $f_{0}$, the pion, kaon and $\eta$ masses
    $M_{\pi},$ $M_{K}$ and $M_{\eta}$, and the octet and decuplet masses $M_{B0}$ and $M_{D0}$,
    respectively. As described in the text, we use two perturbatively equivalent sets for the explicit 
    numerical values entering the fits. }
  \begin{center}
    \begin{tabular}{ccccccccccc}
      \hline 
      \hline 
      appearing quantity &$f_{0}$ {[}MeV{]} & $M_{\pi}$ {[}MeV{]} & $M_{K}$ {[}MeV{]} & $M_{\eta}$ {[}MeV{]} & $M_{B0}$ {[}MeV{]} & $M_{D0}$ {[}MeV{]} & $\mathcal{C}$ & $\mathcal{H}$\tabularnewline
      chiral limit choice &$87$ &  $140$ & $496$ & $547$ & $880$ & $1152$ & $-D$ & $3D-9F$\tabularnewline
      physical average choice   &$1.17\cdot 92$ & $140$ & $496$ & $547$ & $1149$ & $1381$ & $-0.85$ & $3D-9F$\tabularnewline
      \hline 
      \hline 
    \end{tabular}
  \end{center}
\end{table}

\section{Results\label{sec:Results}}

In this section we analyze the SHD data and the lQCD results described in 
Sec.~\ref{sec:semileptonic-hyperon-decays} and listed in 
Tab.~\ref{tab:Semileptonic-hyperon-data}. We use the covariant B$\chi$PT
in the EOMS scheme \cite{Gegelia:1999gf,Fuchs:2003qc} up to $\mathcal{O}(p^3)$, 
which leads to Eq.~(\ref{eq:g1BChPT}) for the octet-baryon AV charges. 
In its complete form, there are eight fitted LECs appearing: $D,$ $F$, $h_{38-41,43,44}$.
The Tab. \ref{tab:Semileptonic-hyperon-data} contains updated 
experimental data as compared to the ones used in previous works. 
We start discussing earlier results obtained in analyses done at LO in the 
chiral expansion or at NLO in the HB$\chi$PT or IR-B$\chi$PT approaches.

\subsection{Leading order results and previous NLO B$\chi$PT analyses
  \label{sub:Leading-order-results}}

We examine first the description of the data at leading order in B$\chi$PT, 
i.e. to $\mathcal{O}(p)$. This is equivalent to the $SU(3)$-symmetric Cabibbo 
model \cite{Cabibbo:1963yz,Cabibbo:2003cu} and the fits are shown in 
Tab. \ref{tab:LO_fits}. 
We approximately 
reproduce the results of \cite{Cabibbo:2003cu}, taking into account
the updated SHD data and also excluding the $\Sigma^{-}\to\Lambda$ channel. Its
consideration worsens the LO fit as it  
produces the highest contribution to the $\chi^2$. However, in the next section
we will see that the description improves at NLO.
For illustration, we also include the lQCD data in one of 
the fits. This gives similarly good results, which already indicates that 
the quark-mass dependence of the AV charges is moderate. 
Additionally, this suggests that the
interpretation given in \cite{Cabibbo:2003cu} that there are only mild 
$SU(3)$ breaking effects in SHD carries also over to unphysical quark masses. 
We will discuss this later in more detail.

\begin{table}[H]
  \caption{\label{tab:LO_fits}Cabibbo model fits to the SHD data. The leading
    order B$\chi$PT is equivalent to the Cabibbo model. In the second
    column we use the old data of \cite{Cabibbo:2003cu} and in the
    others the data of Tab. \ref{tab:Semileptonic-hyperon-data} where
    the $\Xi^{0}\to\Sigma^{+}$ mode is updated \cite{AlaviHarati:2001xk,Batley:2006fc}
    and the $\Sigma^{-}\to\Lambda$ one is included.} 
  
  \begin{center}
    \begin{tabular}{cccc}
      \hline 
      \hline 
      & \cite{Cabibbo:2003cu} & SHD & SHD+lQCD\tabularnewline
      \hline 
      $D$ & $0.804(8)$ & $0.800(8)$ & $0.785(5)$\tabularnewline
      $F$ & $0.463(8)$ & $0.469(8)$ & $0.483(5)$\tabularnewline
      $F/D$ & $0.58$ & $0.59$       & $0.62$\tabularnewline
      $\chi_{\mbox{red}}^{2}$ & $\frac{2.0}{3}=0.7$ & $\frac{13.6}{3}=4.5$ & $\frac{23.5}{17}=1.4$\tabularnewline
      \hline 
      \hline 
    \end{tabular}
  \end{center}
\end{table}

At NLO there are several works on the AV charges in HB$\chi$PT
\cite{Jenkins:1990jv,Jenkins:1991es,Savage:1996zd}.
Many more studies implement a combined chiral and $1/N_c$ expansion that aims at 
exploiting the cancellations between octet and decuplet loop diagrams 
arising in the Large $N_c$ limit~\cite{Dashen:1993as,Dashen:1993jt,Dashen:1994qi,FloresMendieta:2000mz,
FloresMendieta:2012dn,CalleCordon:2012xz}. Here we will restrict ourselves to the 
discussion of the previous analyses of the chiral expansion of $g_1(0)$.

The HB results can be obtained by keeping only the LO term of a 
(non-relativistic) expansion of the covariant loop-functions in powers of 
$1/M_B$. We have checked that we recover all the non-analytic structures 
reported in these earlier HB works, including those concerning the 
decuplet contributions in the SSE~\cite{FloresMendieta:2000mz}. The main difference of
these studies with respect to ours is that all the contributions from  
analytic parts were neglected. That is, 
those stemming from the loop contributions were removed and 
$h_i=0$ was assumed, as well as the approximations 
$\Delta=M_{D0}-M_{B0}=0$ and $M_\pi=0$ were employed.
To account for all this, the errors of the SHD data were increased
globally to $0.2$. In using these same approximations
and the data of Tab. \ref{tab:Semileptonic-hyperon-data},
we are able to qualitatively reproduce the results and conclusions of  
\cite{Jenkins:1990jv,Jenkins:1991es}.

In any case, this treatment of NLO corrections and the increment
of error bars is not systematic.
In particular, we will see below that the role of the analytic 
terms is very important and they are a source for cancellations at NLO that 
dominate those between the octet and the decuplet. 

Apart from the HB studies, the work in IR-B$\chi$PT of Ref.~\cite{Zhu:2000zf}  
correctly incorporated the $\mathcal{O}(p^{3})$ contact
terms, i.e. included all the LECs $h_{i}$. 
To perform fits, the two $SU(3)$ symmetric LECs
$h_{40/41}$ are absorbed into $D$ and $F$, leading to the same number
of fit-parameters as of input data. They specifically investigated the effect of 
the leading recoil corrections in the IR-B$\chi$PT, and found that these, 
which formally are of $\mathcal{O}(p^4)$ in the HB expansion, are typically 
larger than the LO ones. Naturally, the conclusion 
of this study was that the convergence of the chiral expansion of $g_1(0)$ is 
severely broken. 

However, this work has some weaknesses that need to be scrutinized. Firstly, 
the fitted parameters $D$ and $F$ are not those defined in the chiral limit but
an effective parametrization which mixes $\mathcal{O}(p)$ and $\mathcal{O}(p^3)$
contributions. This hinders any definite discussion about the convergence of
the chiral expansion of $g_1(0)$. Secondly, the decuplet contributions are 
neglected despite of the important role they play in reducing the overall size 
of the loop corrections, as suggested by the previous HB and Large $N_c$ studies.
Finally, and most importantly, it must be investigated if the large size of the recoil 
corrections reported in this paper, which are roughly a factor 10 larger than 
expected by power counting, is a genuine problem of the chiral expansion 
or, instead, an artifact introduced by the IR renormalization scheme.
In the next section, we discuss our complete results in the 
EOMS scheme where we tackle all the issues mentioned above.

\subsection{Results in the EOMS-B$\chi$PT at NLO\label{sub:Results p3}}

\begin{figure}
  \caption{\label{fig:Axial-vector-charges}Axial-vector charges $g_1^{BB^\prime}$ 
    and $g_{A,3}^X$ of the SHD and
    octet baryons compared to our fits including explicit virtual decuplet states.
    Blue circle markers denote the fitted input data points. The square markers denote our 
    full B$\chi$PT results while the triangle markers show the LO contribution.
    The shaded area corresponds to the lQCD input from \cite{Lin:2007ap}
    and \cite{Gockeler:2011ze}. In the case of the lQCD data from \cite{Gockeler:2011ze},
    the data points are listed from left to right with increasing $M_{\pi}$. }

  \centering{\includegraphics[scale=0.33]{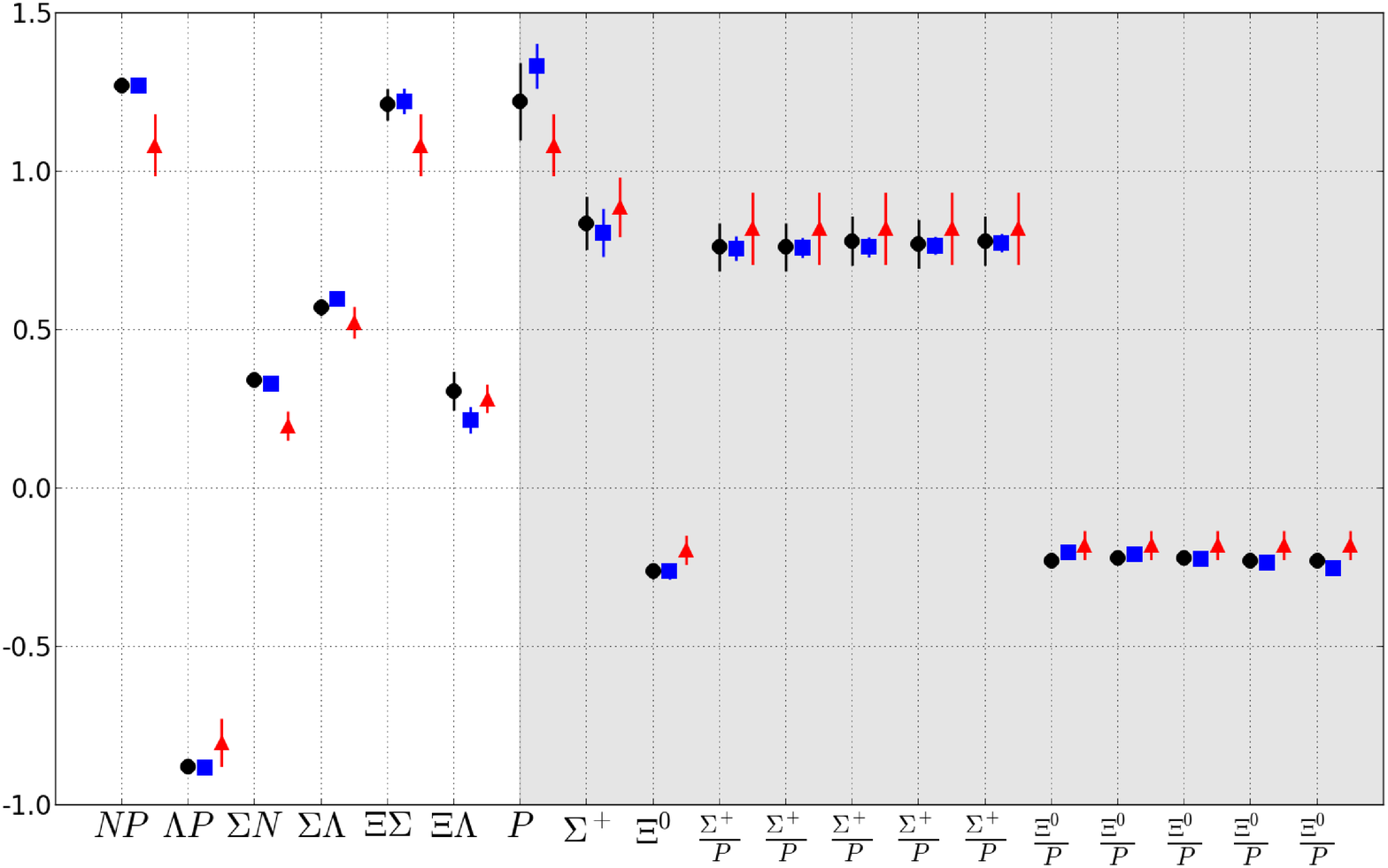}}
\end{figure}

In Tab.~\ref{tab:SHD-contributions} and Fig.~\ref{fig:Axial-vector-charges}, 
we show our final results of the EOMS-B$\chi$PT fits 
to the SHD data and lQCD results discussed in Sec.~\ref{sec:semileptonic-hyperon-decays}.
The description is excellent at NLO whether decuplet 
resonances are explicitly included or not. Also, the 
$\Sigma^- \to \Lambda$ mode can now be consistently described.

For each channel and fit strategy, we separate the different contributions to 
$g_1(0)$. By looking at the last row in Tab.~\ref{tab:Fit-result-paramters}, 
it can be noticed that the total NLO contribution is typically smaller than  
$20\%$ ($30\%$) of the LO one, except 
for the $\Sigma^{-}\to n$ channel which could be up to a $\sim 68\%$ ($\sim118\%$) 
in the theory with (without) explicit decuplets. The overall picture is 
consistent with the naive power counting by which one expects the $\mathcal{O}(p^3)$ 
corrections to be around $(M_k/\Lambda)^2 \sim 25 \%$ of the $\mathcal{O}(p)$ ones. 

As a consequence, B$\chi$PT at NLO is compatible with small 
$SU(3)$-breaking effects in the SHD. This is remarkable since 
only at LO the Lagrangian is fully $SU(3)$ symmetric and most NLO operators 
or loop corrections break the symmetry. This structure, 
together with the actual number of LECs, is not arbitrary but 
dictated by spontaneous chiral symmetry breaking and chiral power counting. In 
practice, the successful description of the SHD data and the good 
convergence are achieved by sizable cancellations between the
different NLO terms. These cancellations are different in the theories with or 
without the explicit decuplet baryons.

In the theory without the decuplet, the loop contributions are given by the 
tadpoles ($T_3$) and the diagrams with internal octet baryons ($B^{loops}$) only.
Individually, these are typically $25\%$ of the LO value although they can be 
as large as a $50\%$. On top of that, they have the same sign in almost all the 
channels. The large $SU(3)$-breaking thus produced is not compatible with the
SHD data and, as a result, the LECs $h_i$ coming from the $\mathcal{O}(p^3)$  
contact-terms ($C_3$) are adjusted in the fit to largely cancel the effects of 
the loops. 

On the other hand, in the theory with explicit decuplet contributions, the new 
loops ($D^{loops}$) can be as sizable as the other ones but, generally, with the 
opposite sign. We observe that the octet-decuplet cancellations found in
\cite{Jenkins:1991es} carry over to the covariant formulation of B$\chi$PT and 
using a finite octet-decuplet mass splitting. The main consequence 
of this is an important reduction of the net contribution of the loops and, hence, 
of the size of the $C_3$ terms. Although the results and overall convergence 
patterns look equivalent in both theories, the inspection of the different 
pieces reveals that the values of the LECs in the theory with the decuplet are
more natural.         

In Tab.~\ref{tab:Fit-result-paramters} we show the values of our fitted
parameters. As discussed in Sec. \ref{BChPT}, one has the freedom at NLO to fix
some the LECs to either their chiral limit values or the average of their 
physical ones. Our default choice is the latter, which corresponds to the results
of Tab. \ref{tab:SHD-contributions}. However, we list the values of the fitted LECs
for both choices and notice that the 
results are rather insensitive with respect to these sets of input parameters. 
We also tested the impact of the decuplet LEC
$\mathcal{H}$ when allowing for a $10 \% \sim 1/N_C^2$ uncertainty to the large-$N_C$
input or fixing it by the $SU(2)$ relation $\mathcal{H}=-g_A(8/5)$.
In all cases we obtain results that are compatible within the statistical 
uncertainties of those given in Tabs. \ref{tab:SHD-contributions} and
\ref{tab:Fit-result-paramters}.

In the last row of Tab.~\ref{tab:Fit-result-paramters} we show the reduced 
$\chi^2$ for the different fits. By comparing them to those from 
the LO fits in Tab.~\ref{tab:LO_fits}, one notes that the description
of the data improves at NLO. The values of $D$ and $F$ change by $\sim 21\%$ 
and $\sim 6\%$, respectively. Furthermore, it is remarkable that at NLO the 
ratio $F/D$ is closer to its Large $N_c$ prediction of 
$2/3$. As for the $\mathcal{O}(p^3)$ LECs, there are large 
differences between the results in the theory with or without the decuplet. 
This is expected on general grounds since the effects of the resonances are 
encoded in the values of the LECs in the latter case.

As a final result for $D$ and $F$ we report
\begin{equation}
  D=0.623(61)(17),\hspace{3cm}F=0.441(47)(2), \label{eq:DandF}
\end{equation}
which is the average between our results with explicit decuplets as listed in 
Tab.~\ref{tab:Fit-result-paramters}. This accounts for the ``naturalness'' 
issue we addressed above for the decuplet-less theory and the fact that 
integrating out decuplet resonances in a $SU(3)$ context is not well justified.
The first error is statistical and the second a systematical one, 
that covers the central values of the two fits. As an interesting 
by-product of our results, we predict the chiral-limit value of 
$g_{A0} = 1.064(77)(19)$ in the $SU(3)$-B$\chi$PT,
which is smaller than the physical AV charge of $g_A = 1.270(3)$.

Having obtained a reliable description of the $g_{1}/f_{1}$ ratios
of the SHD, we are also able to discuss the channels that did not enter
in our fits. These are the SHDs $\Sigma^{-}\to\Sigma^{0}$ and $\Xi^{-}\to\Xi^{0}$
and the isovector AV charges $g_{A,3}^{\Sigma^{_{+}}\Sigma^{_{+}}}$ and $g_{A,3}^{\Xi^{0}\Xi^{0}}$ 
at the physical point.
They are not experimentally measured yet and our values are predictions. 
We list the results in the last four columns of Tab. \ref{tab:SHD-contributions}.
Note that the values shown for the SHD and the charges are related
by isospin. However, for convenience we give both of them explicitly. 

Since we can apply a non-relativistic expansion to our covariant formulas,
we are also able to perform a similar SHD study in the HB formalism. We list 
the results for the decuplet-less case in App. \ref{sec:Heavy-baryon-results}.
Also with this approach, we obtain an excellent description of the SHD data 
with equivalent conclusions to those discussed 
above. These findings, together with our EOMS results above, 
are quite the opposite to those in the covariant IR-B$\chi$PT 
study~\cite{Zhu:2000zf} where very large recoil corrections are reported. 
As a result, we conclude that the stated poor chiral convergence might be related to 
the problems this covariant prescription introduces in the analytic 
structure of the loop functions~\cite{Geng:2008mf,Ledwig:2010nm,Ledwig:2011cx,Alarcon:2012kn}. 
Apart from this, we want to stress that 
the agreement between covariant and HB$\chi$PT 
is quite remarkable given the sizable differences that have been 
found between these approaches in other $SU(3)$-B$\chi$PT applications~\cite{Geng:2008mf,Geng:2009hh,MartinCamalich:2010fp}. 
Probably this is a consequence of the large number of LECs at NLO, 
as it can be seen by comparing the values in the different columns. Differences between
the two approaches might show up in other observables where the values of these 
LECs also appear, e.g. in meson-baryon scattering processes.

\begin{table}[H]
  \caption{\label{tab:SHD-contributions}
    Axial-vector charges and couplings of the octet baryons.
    We used the average of the physical values for the fixed LECs. 
    The results are decomposed into their chiral order contributions, 
    i.e. into LO and the individual
    $p^{3}$ contributions of the graphs $C_{3}$, $T_{3}$ and
    loops with virtual octet baryons $B^{loops}$ or decuplet baryons
    $D^{loops}$. We also show the total NLO contributions relative to
    the LO one. In the last four columns are predictions for the
    strangeness-conserving SHD where the values of $g_{1}^{BB^{\prime}}$ and
    $g_{A,3}^{B}$ are connected by isospin symmetry.}
  \begin{center}
    \begin{tabular}{ccccccccccc}
      \hline 
      \hline 
      $g_{1}$ & $N\to P$ & $\Lambda\to P$ & $\Sigma^{-}\to n$ & $\Sigma^{-}\to\Lambda$ & $\Xi^{0}\to\Sigma^{+}$ & $\Xi^{-}\to\Lambda$ & $\Sigma^{-}\to\Sigma^{0}$ & $\Xi^{-}\to\Xi^{0}$ & $g_{A,3}^{\Sigma^+}$ & $g_{A,3}^{\Xi^0}$\tabularnewline
      \hline 
      Exp & $1.270(3)$ & $-0.879(18)$ & $0.340(17)$ & $0.570(27)$ & $1.210(50)$ & $0.306(61)$ & na & na & na & na \tabularnewline
      \hline 
      Cov       & & & & & & & & & & \tabularnewline
      LO        & $1.16$  & $-0.89$ & $0.15$  & $0.54$  & $1.16$  & $0.35$  & $0.72$  & $0.15$  &  $1.01$ & $-0.150$\tabularnewline
      $C_{3}$    & $-0.45$ & $0.59$  & $0.14$  & $-0.30$ & $-0.92$ & $-0.49$ & $-0.50$ & $-0.01$ & $-0.71$ & $0.01$\tabularnewline
      $T_{3}$    & $0.27$  & $-0.40$ & $0.07$  & $0.13$  & $0.52$  & $0.16$  & $0.17$  & $0.04$  & $0.24$  & $-0.04$\tabularnewline
      $B^{loops}$ & $0.29$  & $-0.19$ & $-0.03$ & $0.23$  & $0.45$  & $0.22$  & $0.34$  & $0.02$  & $0.48$  & $-0.02$\tabularnewline

      full & $1.270(3)$ & $-0.886(18)$ & $0.327(15)$ & $0.597(22)$ & $1.213(38)$ & $0.240(41)$ & $0.718(52)$ & $0.201(38)$ & $1.016(74)$ & $-0.201(38)$\tabularnewline
      $|p^{3}/p^{1}|$ & $0.09$ & $\sim 0.0$ & $1.18$ & $0.11$ & $0.04$ & $0.32$ & $\sim0$ & $0.33$ & $\sim0$ & $0.33$\tabularnewline
      \hline 
      Cov+D     & & & & & & & & & &\tabularnewline
      LO        & $1.08$  & $-0.80$ & $0.20$  & $0.52$  & $1.08$  & $0.28$  & $0.62$  & $0.20$  & $0.89$  & $-0.20$\tabularnewline
      $C_{3}$    & $-0.18$ & $0.24$  & $0.01$  & $-0.18$ & $-0.28$ & $-0.17$ & $0.08$  & $-0.22$ & $0.11$  & $0.22$\tabularnewline
      $T_{3}$    & $0.25$  & $-0.36$ & $0.09$  & $0.12$  & $0.48$  & $0.13$  & $0.15$  & $0.05$  & $0.21$  & $-0.05$\tabularnewline
      $B^{loops}$ & $0.18$  & $-0.09$ & $-0.02$ & $0.17$  & $0.30$  & $0.12$  & $0.24$  & $0.06$  & $0.34$  & $-0.06$\tabularnewline
      $D^{loops}$ & $-0.07$ & $0.13$  & $0.06$  & $-0.03$ & $-0.37$ & $-0.15$ & $-0.36$ & $0.15$  & $-0.51$ & $-0.15$\tabularnewline

      full      & $1.270(3)$ & $-0.883(18)$ & $0.330(16)$ & $0.597(22)$ & $1.221(40)$ & $0.214(42)$ & $0.740(55)$ & $0.225(35)$ & $1.047(77)$ & $-0.225(35)$\tabularnewline
      $|p^{3}/p^{1}|$ & $0.17$ & $0.10$ & $0.68$ & $0.14$ & $0.13$ & $0.24$ & $0.18$ & $0.15$ & $0.18$ & $0.15$\tabularnewline
      \hline 
      \hline 
    \end{tabular}
  \end{center}
\end{table}

\begin{table}[H]
  \caption{\label{tab:Fit-result-paramters}
    Fit results of our EOMS B$\chi$PT analysis of the data in 
    Tab. \ref{tab:Semileptonic-hyperon-data}.
    In the large $N_{C}$ limit one has $F/D=2/3$ and $\mathcal{C}=-D$. We 
    list the results with respect to the choices of fixed parameters as
    shown in Tab. \ref{tab:Fit-parameter}. The choice of chiral limit
    parameters is marked with \it{chiral}.
  }

\begin{center}
  \begin{tabular}{ccccc}
    \hline 
    \hline 
                        & Cov           & Cov+D         & Cov \it{chiral} & Cov+D \it{chiral} \tabularnewline
    \hline 
    $D$                 & $0.658(64)$   & $0.639(61)$   & $0.634(59)$   & $0.606(53)$    \tabularnewline
    $F$                 & $0.507(56)$   & $0.443(47)$   & $0.492(52)$   & $0.439(42)$    \tabularnewline
    $F/D$               & $0.77$        & $0.69$        & $0.78$        & $0.72$         \tabularnewline
    $h_{38}$ [GeV$^{-2}$] & $0.146(29)$   & $-0.008(34)$  & $0.143(27)$   & $-0.051(37)$   \tabularnewline
    $h_{39}$ [GeV$^{-2}$] & $-0.002(36)$  & $0.032(36)$   & $0.006(38)$   & $0.049(42)$    \tabularnewline
    $h_{40}$ [GeV$^{-2}$] & $-0.349(135)$ & $-0.077(102)$ & $-0.354(133)$ & $-0.030(52)$   \tabularnewline
    $h_{41}$ [GeV$^{-2}$] & $-0.009(47)$  & $-0.151(49)$  & $-0.004(45)$  & $-0.113(59)$   \tabularnewline
    $h_{43}$ [GeV$^{-2}$] & $0.082(39)$   & $0.159(49)$   & $0.083(39)$   & $0.159(53)$    \tabularnewline
    $h_{44}$ [GeV$^{-2}$] & $-0.111(26)$  & $-0.062(22)$  & $-0.118(25)$  & $-0.062(15)$   \tabularnewline

    $\chi_{\mbox{red}}^{2}$ & $\frac{7.0}{11}=0.64$ & $\frac{7.4}{11}=0.67$ & $\frac{7.2}{11}=0.65$ & $\frac{7.9}{11}=0.72$\tabularnewline
    \hline 
    \hline 
  \end{tabular}
\end{center}
\end{table}

\begin{figure}
  \caption{\label{fig:Chiral-extrapolation}Chiral extrapolation of the AV
    charges $g_{A,3}^X$ of the proton (red), $\Sigma^+$ (blue)
    and $\Xi^0$ (green) as function of $M_\pi$.
    The left figure shows the lowest $M_{\pi}$ data from \cite{Lin:2007ap}
    with $M_{k}=604$ MeV. The right figure shows the data from \cite{Gockeler:2011ze}
    along the $SU(3)$ singlet line for $g_{A,3}^{\Sigma^+}/g_{A,3}^{P}$ (blue) 
    and $g_{A,3}^{\Xi^0}/g_{A,3}^{P}$ (green). We also plot the LO 
    contribution as obtained from the full $p^{3}$ fit, dashed line, as well as the LO
    results of Sec. \ref{sub:Leading-order-results}, dotted line.}
    \begin{center}
      \includegraphics[scale=0.43]{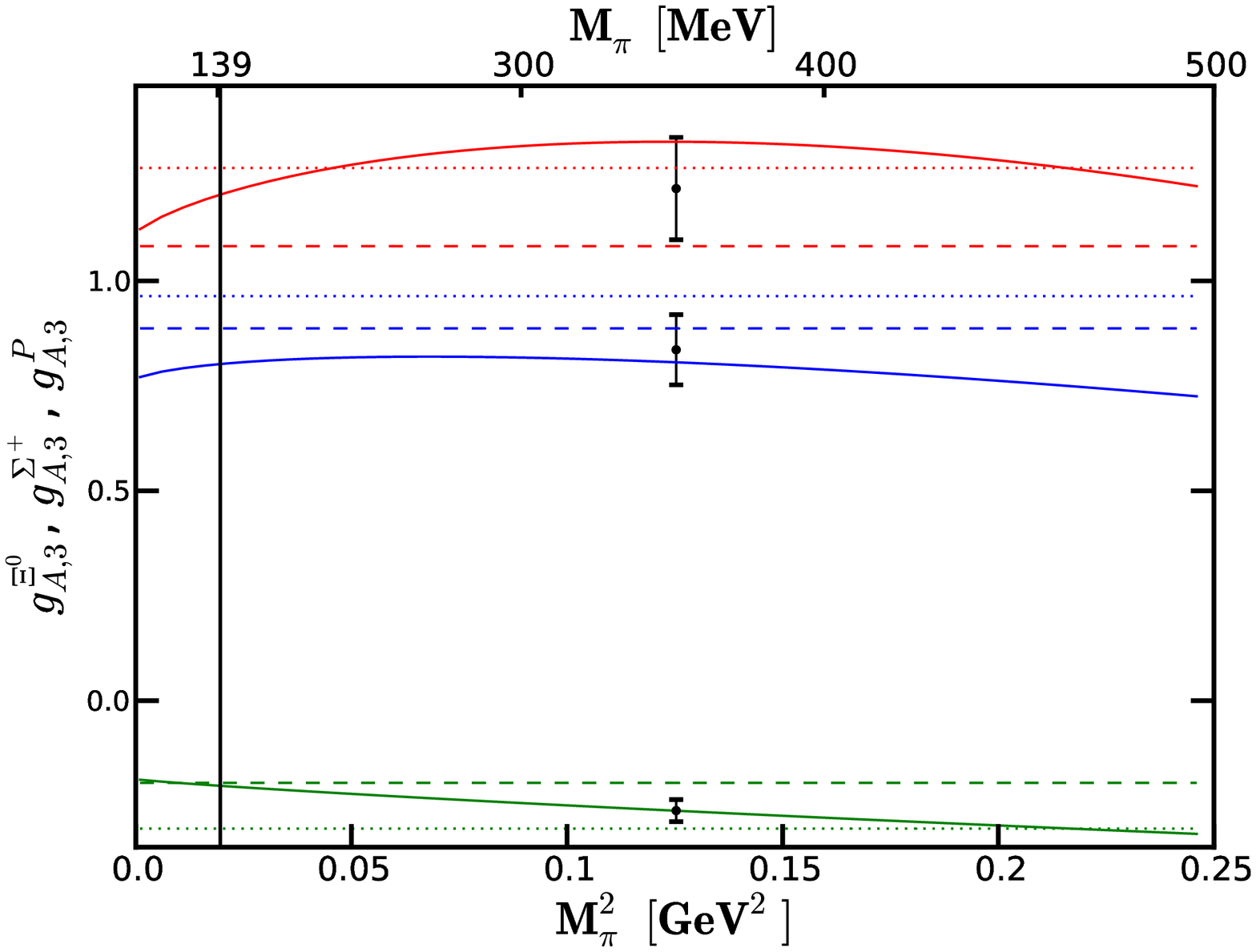}
      \includegraphics[scale=0.43]{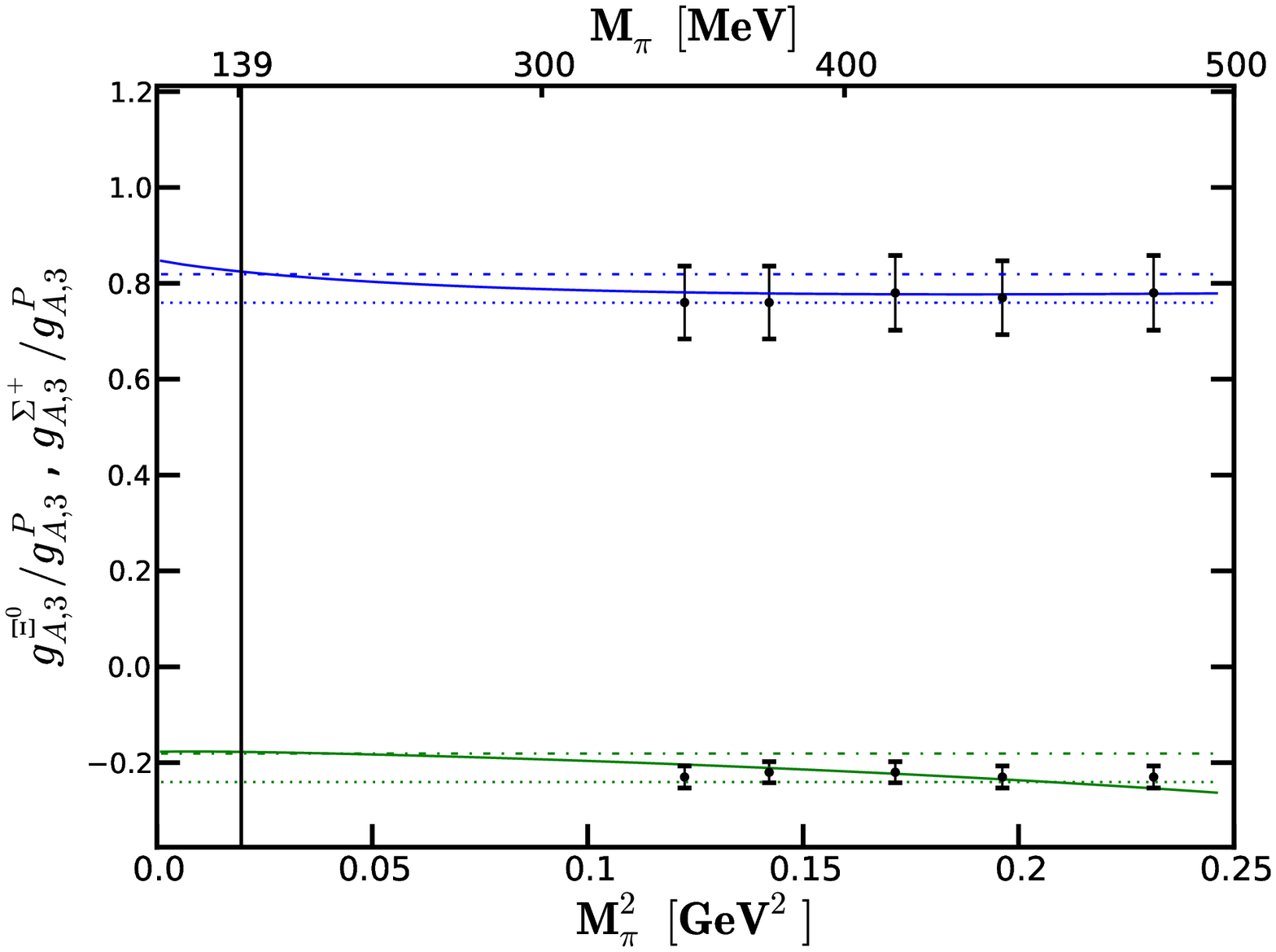}
    \end{center}
\end{figure}

\begin{figure}
  \caption{\label{fig:Chiral-convergence} Chiral convergence of the AV
    charges. We show the ratio of NLO contributions over the LO one for the 
    $g_{A,3}^X$ of the proton (red), $\Sigma^+$ (blue) and $\Xi^0$ (green) as function of $M_\pi$.}
    \begin{center}
      \includegraphics[scale=0.43]{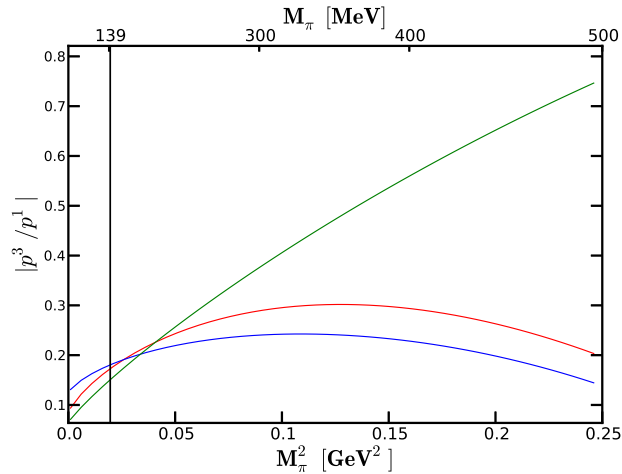}
    \end{center}
\end{figure}

Finally, we are also able to discuss how the $SU(3)$-breaking effects behave 
for unphysical quark-masses. In Figs. \ref{fig:Chiral-extrapolation} and 
\ref{fig:Chiral-convergence} we show the chiral behavior of the isovector 
AV charges as function of $M_{\pi}$, 
together with the ratios of the NLO contributions over the LO ones, i.e.
their chiral convergences. We also plot the LO contributions of the NLO fits 
as well as the results of the pure Cabibbo model fits (LO B$\chi$PT) of 
Sec. \ref{tab:LO_fits}. 

We see that the chiral behavior is quite flat and is in very good agreement
with the LO result and the dependence shown by the lQCD studies. 
Therefore, the cancellations among various $p^{3}$ terms at the physical point
also hold for unphysical quark masses. The overall chiral
convergence is very acceptable for the whole quark-mass region. 
Similar chiral extrapolations can be found in the theory without explicit 
decuplet states as well as in the HB approach.

\subsection{Octet axial-vector charges and the quark contribution to the proton's spin
\label{sec:proton spin} }

A very important application of the study of the SHDs has been the prediction 
of the octet axial charge of the proton, $g_A^{8}$. This is
defined as the axial charge corresponding to 
$\langle P(p^\prime)| \overline{q}\gamma^\mu\gamma_5\lambda^8 q| P(p^\prime)\rangle$ 
and its importance lies on the fact that it gives a crucial constraint to obtain
 the flavor structure of the quark contribution to the proton's spin (see~\cite{Aidala:2012mv}
 for a recent review). Even though this is an old and persisting question in 
nucleon structure, the value of $g_A^8$ is not well known yet. At LO in the 
chiral expansion one recovers the $SU(3)$ prediction that $g_{A}^8=3F-D\simeq0.58$.
 The success of the Cabibbo model in the description of the SHDs suggests that 
this determination could be accurate and it is often used in the phenomenological
 analyses. However, a model-independent understanding of the quark contribution
 to the proton spin requires a better determination of $g_A^8$ and efforts in this
 direction have been undertaken in lQCD~\cite{QCDSF:2011aa,Engelhardt:2012gd}.
 \footnote{As a side remark,
 we note that the quark contribution to the proton spin is an important input 
parameter for constraining BSM parameters from the spin-dependent cross-section 
in direct dark matter searches.}

In principle, B$\chi$PT can be used to improve the determination to higher
orders in the chiral expansion. Such studies were carried out in the HB~\cite{Jenkins:1990jv} 
and IR~\cite{Zhu:2000zf} schemes and the conclusions were in both cases that the NLO correction 
could be very large which hampered the convergence of the chiral expansion of $g_A^8$.
However, these conclusions are afflicted by the same caveats as those addressed above
in Sec.~\ref{sub:Leading-order-results}, and they should be revised in the context 
of the current full NLO calculation. Furthermore, the octet axial charge receives a 
contribution from a LEC, $h_{42}$, that is not constrained by SHDs data as it does not 
contribute to the flavor-changing transitions 
or to the isospin related isovectorial charges. This fact has been overlooked in
the previous chiral analyses and it precludes a determination of $g_A^8$ from SHDs alone.

\begin{table}[H]
  \caption{\label{tab:g8}Different contributions to $g_A^8$ in EOMS-B$\chi$PT up to 
    $\mathcal{O}(p^3)$ and assuming $g_A^8=0.58$.}
  
  \begin{center}
    \begin{tabular}{ccccccc}
      \hline  \hline 
      &LO&NLO&$C_3$&$T_3$&$B_3$&$D_3$\\
      \hline
      Octet &0.84&$-0.26$&$-0.72$&0.55&$-0.09$&$-$\\
      Octet+Decuplet &0.71&$-0.13$&$-0.38$&0.47&$-0.13$&$-0.09$\\
      \hline \hline 
    \end{tabular}
\end{center}
\end{table}

Nevertheless, even without a precise value for $h_{42}$ we are able to study the 
convergence of $g_A^8$ under quite general assumptions. For
this, we assume that $g_A^8$ is close to its $SU(3)$-symmetric
value, $g_A^8\sim0.58$, as suggested by a recent lQCD 
determination, and we fix $h_{42}$. The size of 
the different contributions up to $\mathcal{O}(p^3)$ are 
shown in Tab.~\ref{tab:g8}. By comparing the overall LO and 
NLO contributions, we see that the 
convergence in this scenario is good, with NLO corrections
about a 20\% (30\%) the LO ones in the decuplet (decuplet-less)
theory. In the theory without decuplets one finds that 
the total loop contribution is quite large. As a consequence,
the NLO contact-terms are sizable and as large as the total 
LO. This leads to the same naturalness considerations discussed
above for the AV charges. On the other hand, the diagrams with
decuplet baryons reduce the net loop contribution and
improve the convergence.     

The current analysis clarifies the structure of the chiral 
expansion of the octet axial coupling of the proton and opens 
the possibility for a model-independent treatment of its $SU(3)$-breaking 
corrections and for an improvement of the phenomenological 
extractions of the quark content of the proton's spin. 

\section{Summary}

We have studied the axial-vector charges of the octet baryons in $SU(3)$ 
covariant B$\chi$PT up to $\mathcal{O}(p^3)$ using the EOMS scheme and
including decuplet resonances. We report that B$\chi$PT at this 
next-to-leading order consistently describes the charges as well as the ratios
$g_1/f_1$ of the axial-vector and vector couplings measured in the 
semileptonic hyperon decays. This is a novel feature as compared 
with previous B$\chi$PT studies in the non-relativistic heavy-baryon scheme or the 
relativistic infrared approach.

Explicitly, we have been able to determine all appearing low-energy constants
from simultaneous fits to the semileptonic hyperon decay data and
available lQCD results. This includes the leading-order constants $D$ and 
$F$ as well as all the NLO constants $h_{38-41,43,44}$. Along this, 
we have clarified the role of the different contact-terms appearing at this order 
which were not treated systematically in the previous works. 
Especially, we disentangle the two singlet LECs $h_{40,41}$ from $D$ and $F$, 
which lead to an accurate 
determination of the latter and a consistent discussion of the chiral 
convergence for the axial-vector charges. 

We report a systematic improvement of the theoretical understanding of the data 
with respect to the $SU(3)$-symmetric Cabibbo model, which is equivalent
to the B$\chi$PT at LO. That is, at NLO we are also able to consistently 
include the mode $\Sigma^- \to \Lambda$, as well as we obtain NLO 
corrections that are typically 20\% of the LO ones. This size of NLO
effects is in agreement with the naive power counting. Therefore, our analysis
shows that $SU(3)$-symmetry-breaking effects, as given by the spontaneous chiral 
symmetry and the chiral power counting in B$\chi$PT, are 
important to understand the SHD data accurately.

In practice, the agreement at $\mathcal{O}(p^3)$ is achieved by sizable 
cancellations between different $SU(3)$ breaking terms, in particular 
those parameterized by the NLO LECs $h_i$ and the ones from the loops. 
We showed that considering only NLO non-analytic terms is not 
enough and that NLO analytic terms play an important role. 
The cancellations themselves appear in both 
theories with and without explicit decuplet
states, however, they have a different structure.
In the case with decuplet baryons, we found that 
their explicit contributions are a source of 
cancellations which lead to more natural values for the NLO LECs. 
This is in agreement with the expectations derived from the analysis at 
large $N_c$. Furthermore, the fact that in the decuplet-theory we can 
successfully describe the small $SU(3)$-breaking in $g_1(0)$ by means of a
chiral expansion without anomalously large or small chiral corrections at 
NLO is a very non-trivial outcome of our study.

A phenomenological consequence of our work is the determination of the LO
axial couplings $D=0.623(61)(17)$ and $F=0.441(47)(2)$ up to $\mathcal{O}(p^3)$ 
accuracy in a completely systematic fashion. Remarkably, these values are 
closer to the Large $N_c$ ratio $D/F\sim2/3$ than at LO and they predict
the axial coupling of the nucleon in the chiral limit and in $SU(3)$ to be 
$g_{A0}=1.064(77)(19)$. We also predict the isovector axial-vector charges for the
$\Lambda$, $\Sigma^{+}$ and $\Xi^{0}$ or, equivalently, for the SHD channels
of $\Sigma^{-}\to\Sigma^{0}$ and $\Xi^{-}\to\Xi^{-}$.

Finally, we have discussed an important application of the
analysis of the axial-vector charges, namely the prediction of the octet axial coupling
$g_A^8$ and its role in the determination of the quark content of the proton 
spin. More specifically, we have found that there is a contribution from 
a NLO contact-term (whose LEC is labelled by $h_{42}$) which is unconstrained
by SHDs. Therefore one needs additional experimental or nonperturbative
information to determine this parameter. Nevertheless, we studied the 
chiral convergence of $g_A^8$ and concluded that it is reasonable. This 
should allow for a model independent determination of this quantity.

\begin{acknowledgments}
This work has been supported by the Spanish Ministerio
de Economía y Competitividad and European FEDER funds under Contracts
FIS2011-28853-C02-01, Generalitat Valenciana
under contract PROMETEO/2009/0090 and the EU Hadron-Physics3 project,
Grant No. 283286. J.M.C has received funding from the People Programme
(Marie Curie Actions) of the European Union's Seventh Framework Programme
(FP7/2007-2013) under REA grant agreement n PIOF-GA-2012-330458. LSG was 
partly supported by the National Natural Science Foundation of China under
Grant No. 11375024 and  the New Century Excellent Talents in University
Program of Ministry of Education of China under Grant No. NCET-10-0029.
\end{acknowledgments}
\begin{appendix}

\section{Notation\label{App:Notation}}

The notation for the baryon Lagrangian Eq. (\ref{eq:Lfull}) is as
follows. \\
The meson field $\phi=\phi(x)$ is defined by 
\begin{eqnarray}
  U & = & e^{i\frac{\sqrt{2}}{f_{0}}\phi}\,\,\,\,\mbox{and}\,\,\,\, u=\sqrt{U}\\
  \phi & = & \frac{\lambda^{a}\phi^{a}}{\sqrt{2}}=\frac{1}{\sqrt{2}}\left[\begin{array}{ccc}
      \phi^{3}+\frac{1}{\sqrt{3}}\phi^{8} & \phi^{1}-i\phi^{2} & \phi^{4}-i\phi^{5}\\
      \phi^{1}+i\phi^{2} & -\phi^{3}+\frac{1}{\sqrt{3}}\phi^{8} & \phi^{6}-i\phi^{7}\\
      \phi^{4}+i\phi^{5} & \phi^{6}+i\phi^{7} & -\frac{2}{\sqrt{3}}\phi^{8}
    \end{array}\right]=\left[\begin{array}{ccc}
      \frac{1}{\sqrt{2}}\pi^{0}+\frac{1}{\sqrt{6}}\eta & \pi^{+} & K^{+}\\
      \pi^{-} & -\frac{1}{\sqrt{2}}\pi^{0}+\frac{1}{\sqrt{6}}\eta & K^{0}\\
      K^{-} & \overline{K}^{0} & -\frac{2}{\sqrt{6}}\eta
    \end{array}\right]\,\,\,.
\end{eqnarray}
The octet baryon field $B\left(x\right)$ is defined by

\begin{eqnarray}
  B=\frac{\lambda_{a}}{\sqrt{2}}B_{a} & = & \left[\begin{array}{ccc}
      \frac{1}{\sqrt{2}}\Sigma^{0}+\frac{1}{\sqrt{6}}\Lambda & \Sigma^{+} & p\\
      \Sigma^{-} & -\frac{1}{\sqrt{2}}\Sigma^{0}+\frac{1}{\sqrt{6}}\Lambda & n\\
      \Xi^{-} & \Xi^{0} & -\frac{2}{\sqrt{6}}\Lambda
    \end{array}\right]=\left(B^{ab}\right)\,\,\,\,,\\
  \overline{B}=\frac{\lambda_{a}}{\sqrt{2}}\overline{B}_{a} & = & \left[\begin{array}{ccc}
      \frac{1}{\sqrt{2}}\overline{\Sigma}^{0}+\frac{1}{\sqrt{6}}\overline{\Lambda} & \overline{\Sigma}^{-} & \overline{\Xi}^{-}\\
      \overline{\Sigma}^{+} & -\frac{1}{\sqrt{2}}\overline{\Sigma}^{0}+\frac{1}{\sqrt{6}}\overline{\Lambda} & \overline{\Xi}^{0}\\
      \overline{p} & \overline{n} & -\frac{2}{\sqrt{6}}\overline{\Lambda}
    \end{array}\right]=\left(\overline{B}^{ab}\right)\,\,\,\,.
\end{eqnarray}
The decuplet field $T\left(x\right)$ is defined by the totally symmetric
tensor $T^{abc}$ 

\begin{eqnarray}
  T_{111}^{\alpha}=\Delta^{++\alpha}\,\,\,\, T_{112}^{\nu}=\frac{1}{\sqrt{3}}\Delta^{+\alpha} & \,\,\,\, & T_{122}^{\alpha}=\frac{1}{\sqrt{3}}\Delta^{0\alpha}\,\,\,\, T_{222}^{\alpha}=\Delta^{-\alpha}\\
  T_{113}^{\nu}=\frac{1}{\sqrt{3}}\Sigma^{*+\alpha} & \,\,\,\, T_{123}^{\nu}=\frac{1}{\sqrt{6}}\Sigma^{*0\alpha}\,\,\,\, & T_{223}^{\alpha}=\frac{1}{\sqrt{3}}\Sigma^{*-\alpha}\\
T_{133}^{\nu}=\frac{1}{\sqrt{3}}\Xi^{*0\alpha} & \,\,\,\, & T_{233}^{\alpha}=\frac{1}{\sqrt{3}}\Xi^{*-\alpha}\,\,\,\,\\
& T_{333}^{\alpha}=\Omega^{-\alpha} & \,\,\,\,,
\end{eqnarray}
together with the decuplet baryon propagator as

\begin{flushleft}
  \begin{equation}
    S_{\Delta}^{\alpha\beta}(p)=\frac{\s p+M_{D0}}{p^{2}-M_{D0}^{2}+i\varepsilon}\left[-g^{\alpha\beta}+\frac{1}{D-1}\gamma^{\alpha}\gamma^{\beta}+\frac{1}{(D-1)M_{D0}}(\gamma^{\alpha}p^{\beta}-\gamma^{\beta}p^{\alpha})+\frac{D-2}{(D-1)M_{D0}^{2}}p^{\alpha}p^{\beta}\right]\,\,\,.
  \end{equation}
  The external axial-vector field is defined by
\begin{equation}
  a_{\mu}=a_{\mu}^{a}\left(x\right)\frac{\lambda^{a}}{\sqrt{2}}=\left[\begin{array}{ccc}
      \frac{1}{\sqrt{2}}a^{0}+\frac{1}{\sqrt{6}}a^{\eta} & a^{\pi^{+}} & a^{K^{+}}\\
      a^{\pi^{-}} & -\frac{1}{\sqrt{2}}a^{0}+\frac{1}{\sqrt{6}}a^{\eta} & a^{K^{0}}\\
      a^{K^{-}} & a^{\overline{K}^{0}} & -\frac{2}{\sqrt{6}}a^{\eta}
    \end{array}\right]\,\,\,\,.
\end{equation}
All other $\chi$PT quantities appearing in Eq. (\ref{eq:Lfull})
are given by
\par\end{flushleft}

\begin{eqnarray}
  \Gamma_{\mu} & = & \frac{1}{2}\left[u^{\dagger}\left(\partial_{\mu}u\right)+u\left(\partial_{\mu}u^{\dagger}\right)\right]-\frac{i}{2}\left[u^{\dagger}r_{\mu}u+ul_{\mu}u^{\dagger}\right]\,\,\,,\\
  u_{\mu} & = & i\left[u^{\dagger}\left(\partial_{\mu}u\right)-u\left(\partial_{\mu}u^{\dagger}\right)\right]+\left[u^{\dagger}r_{\mu}u-ul_{\mu}u^{\dagger}\right]\,\,\,,\\
  \chi_{+} & = & u^{\dagger}\chi u^{\dagger}+u\chi^{\dagger}u\\
  D_{\mu}B & = & \partial_{\mu}B+\left[\Gamma_{\mu},B\right]\\
  \chi & = & 2B_{0}\mbox{diag}\left[\overline{m},\overline{m},m_{s}\right]=\mbox{diag}\left[m_{\pi}^{2},m_{\pi}^{2},2m_{K}^{2}-m_{\pi}^{2}\right]\,\,\,.\\
  f_{-}^{\mu\nu} & = & uF_{L}^{\mu\nu}u^{\dagger}-u^{\dagger}F_{R}^{\mu\nu}u\\
F_{X}^{\mu\nu} & = & \partial^{\mu}X^{\nu}-\partial^{\nu}X^{\mu}-i\left[X^{\mu},X^{\nu}\right]\,\,\,\mbox{with}\,\,\,\, X=r,l
\end{eqnarray}
 The external fields $r_{\mu}=v_{\mu}+a_{\mu}$ and $l_{\mu}=v_{\mu}-a_{\mu}$
contain the external vector and axial-vector fields $v_{\mu}=v_{\mu}\left(x\right)$
and $a_{\mu}=a_{\mu}\left(x\right)$. For the present work we set
$v_{\mu}=0$.

The Lagrangian Eq. (\ref{eq:Lfull}) produces the contributions of
Fig. \ref{fig: g1 diagrams} for which we need the following loop
integrals in $D=4-2\varepsilon$ dimensions
\begin{eqnarray}
J_{0}\left(\mathcal{M}^{2},\Lambda^{2}\right) & = & \frac{-i}{\left(4\pi\right)^{2}}\,\left(\mathcal{M}^{2}\right)^{2}\left[L-1+\ln\frac{\mathcal{M}^{2}}{\Lambda^{2}}-\frac{1}{2}\right]\,\,\,\,,\\
J_{1}\left(\mathcal{M}^{2},\Lambda^{2}\right) & = & \frac{-i}{\left(4\pi\right)^{2}}\,\mathcal{M}^{2}\,\left[L-1+\ln\frac{\mathcal{M}^{2}}{\Lambda^{2}}\right]\,\,\,\,,\\
J_{2}\left(\mathcal{M}^{2},\Lambda^{2}\right) & = & \frac{-i}{\left(4\pi\right)^{2}}\,\left[L+\ln\frac{\mathcal{M}^{2}}{\Lambda^{2}}\right]\,\,\,\,,\\
J_{3}\left(\mathcal{M}^{2},\Lambda^{2}\right) & = & \frac{-i}{\left(4\pi\right)^{2}}\,\frac{1}{2}\,\frac{1}{\mathcal{M}^{2}}\,\,\,\,,
\end{eqnarray}
with $L=-\frac{1}{\varepsilon}+\gamma_{E}-\ln4\pi$. We renormalize
all contributions proportional to $L$. 

\section{Axial-vector form factors\label{App:Axial-vector-form-factors}}

We list here all unrenormalized results of Figs. \ref{fig: g1 diagrams}
and \ref{fig: self-energies} that contribute to the structure $\overline{u}^{\prime}\left(p^{\prime}\right)\gamma^{\mu}\gamma_{5}u\left(p\right)$
at $q^{2}=0$. The explicit contributions of Eq. (\ref{eq:g1BChPT})
for a given process $B^{\prime}\to B$ are: 
\begin{eqnarray}
C_{1}=K_{C1}\,\,\,\,,\\
C_{3}=K_{C3}\,\,\,\,,\\
X=\sum_{\phi=\pi,K,\eta}X^{\phi} &  & \mbox{for}\,\,\,\, X=T_{3},B_{3ab},B_{3},D_{3ab},D_{3},\Sigma_{B3}\left(\s p\right),\Sigma_{D3}\left(\s p\right)\,\,\,\,,\\
\sqrt{Z_{B}} & = & 1+\frac{1}{2}\frac{\partial}{\partial\s p}\Sigma_{B}^{\left(3\right)}\left(M_{B0}\right)+\mathcal{O}\left(p^{3}\right)
\end{eqnarray}
with

\begin{eqnarray}
T_{3}^{\phi} & = & -K_{T3}^{\phi}\frac{M_{\phi}^{2}}{\left(4\pi f_{0}\right)^{2}}\left[L-1+\ln\frac{M_{\phi}^{2}}{\Lambda^{2}}\right]\,\,\,\,,\\
B_{3ab}^{\phi} & = & -i\left[K_{3a}^{\phi}-K_{3b}^{\phi}\right]\frac{1}{f_{0}^{2}}\int_{0}^{1}dz\left[\left(-2+\varepsilon\right)J_{1}^{B}-M_{B}^{2}z^{2}J_{2}^{B}\right]\,\,\,\,,\\
B_{3}^{\phi} & = & iK_{3}^{\phi}\frac{1}{f_{0}^{2}}\int_{0}^{1}dz2z\,\left[\left(3-\frac{5}{2}\varepsilon\right)J_{1}^{B}-M_{B}^{2}\left(-1-3z^{2}+\varepsilon\left(2+z^{2}\right)\right)J_{2}^{B}+M_{B}^{4}z^{4}J_{3}^{B}\right]\,\,\,\,,\\
D_{3ab}^{\phi} & = & -i\left[K_{D3a}^{\phi}+K_{D3b}^{\phi}\right]\frac{\mathcal{C}^{2}}{f_{0}^{2}M_{D0}^{2}}\int_{0}^{1}dy\int_{0}^{1}dz2z\,\frac{1}{18}M_{B}\\
 &  & \left(((9-6\varepsilon)M_{D}+M_{B}(30-12z+\varepsilon(7z-19)))J_{1}^{DB}+(\varepsilon-3)M_{B}^{2}(z-2)^{2}(\left(z-1\right)M_{B}-M_{D})J_{2}^{DB}\right)\,\,\,\,,\nonumber \\
D_{3}^{\phi} & = & iK_{D3}^{\phi}\frac{\mathcal{C}^{2}\mathcal{H}}{f_{0}^{2}M_{D}^{4}}\int_{0}^{1}dz2z\,\,\,\frac{1}{108}M_{0}^{2}\Big[\nonumber \\
 &  & -2\left((109\varepsilon-60)M_{D}^{2}-2(139\varepsilon-60)M_{B}M_{D}(z-1)+6(41\varepsilon-15)M_{B}^{2}(z-1)^{2}\right)J_{1}^{D}\nonumber \\
 &  & +(60-319\varepsilon)J_{0}^{D}-2(31\varepsilon-15)M_{B}^{2}(1-z)^{2}(M_{B}+M_{D}-M_{B}z)^{2}J_{2}^{D}\Big]\,\,\,\,,\\
\Sigma_{B3}^{\phi}\left(\s p\right) & = & iK_{B}^{\phi}\frac{1}{f_{0}^{2}}\int_{0}^{1}dz\left[z^{2}p^{2}\left(\s p\left(1-z\right)-M_{B}\right)J_{2}^{B}+\left(\left(-2+\varepsilon\right)M_{B}-\left(1+3z\right)\s p+\varepsilon\left(1+z\right)\s p\right)J_{1}^{B}\right]\,\,\,,\\
\Sigma_{D3}^{\phi}\left(\s p\right) & = & iK_{D}^{\phi}\frac{\mathcal{C}^{2}}{f_{0}^{2}M_{D}^{2}}\int_{0}^{1}dz\s p^{2}\left(z\s p+M_{D}\right)\left(1-\varepsilon\right)J_{1}^{D}\,\,\,\,,
\end{eqnarray}
with $J_{i}^{X}=J_{i}^{X}\left(\mathcal{M}_{X}^{2},\Lambda^{2}\right)$
and 
\begin{eqnarray}
  \mathcal{M}_{B}^{2} & = & \left(1-z\right)M_{\phi}^{2}-z\left(1-z\right)p^{2}+zM_{B}^{2}\,\,\,\,,\\
  \mathcal{M}_{DB}^{2} & = & \left(1-z\right)M_{\phi}^{2}+z^{2}M_{B}^{2}+zy\left(M_{D}^{2}-M_{B}^{2}\right)\,\,\,\,,\\
  \mathcal{M}_{D}^{2} & = & \left(1-z\right)M_{\phi}^{2}-z\left(1-z\right)p^{2}+zM_{D}^{2}\,\,\,\,.
\end{eqnarray}
All the coefficients $K_{i}$ are listed in the tables \ref{tab:CoeffSHD}-\ref{tab:CoeffSelfEnergies}.

\begin{table}[H]
  \caption{\label{tab:CoeffSHD} Coefficients of the graphs $C_{1}$, $T_{3}$,
    $B_{3ab}$, $B_{3}$ $C_{3}$ and the decuplet ones $D_{3ab}$ and
    $D_{3}$ contributing to the semileptonic hyperon decays.}
  
  \begin{center}
    \begin{tabular}{cccccccc}
      \hline 
      \hline 
      & $n\to p$ & $\Lambda\to p$ & $\Sigma^{-}\to n$ & $\Sigma^{-}\to\Lambda$ & $\Xi^{0}\to\Sigma^{+}$ & $\Xi^{-}\to\Sigma^{0}$ & $\Xi^{-}\to\Lambda$\tabularnewline
      \hline 
      $K_{C1}$ & $D+F$ & $-\sqrt{\frac{1}{6}}\left(D+3F\right)$ & $D-F$ & $\sqrt{\frac{2}{3}}D$ & $D+F$ & $\sqrt{\frac{1}{2}}\left(D+F\right)$ & $\sqrt{\frac{1}{6}}\left(3F-D\right)$\tabularnewline
      \hline 
      $K_{T3}^{\pi}$ & $D+F$ & $-\frac{1}{8}\sqrt{\frac{3}{2}}\left(D+3F\right)$ & $\frac{3}{8}\left(D-F\right)$ & $\sqrt{\frac{2}{3}}D$ & $\frac{3}{8}\left(D+F\right)$ & $\frac{3}{8\sqrt{2}}\left(D+F\right)$ & $-\frac{1}{8}\sqrt{\frac{3}{2}}\left(D-3F\right)$\tabularnewline
      $K_{T3}^{K}$ & $\frac{1}{2}\left(D+F\right)$ & $-\frac{1}{4}\sqrt{\frac{3}{2}}\left(D+3F\right)$ & $\frac{3}{4}\left(D-F\right)$ & $\sqrt{\frac{1}{6}}D$ & $\frac{3}{4}\left(D+F\right)$ & $\frac{3}{4}\sqrt{\frac{1}{2}}\left(D+F\right)$ & $-\frac{1}{4}\sqrt{\frac{3}{2}}\left(D-3F\right)$\tabularnewline
      $K_{T3}^{\eta}$ & $0$ & $-\frac{1}{8}\sqrt{\frac{3}{2}}\left(D+3F\right)$ & $\frac{3}{8}\left(D-F\right)$ & $0$ & $\frac{3}{8}\left(D+F\right)$ & $\frac{3}{8\sqrt{2}}\left(D+F\right)$ & $-\frac{1}{8}\sqrt{\frac{3}{2}}\left(D-3F\right)$\tabularnewline
      \hline 
      $K_{3a}^{\pi}$ & $D+F$ & $-\frac{3}{4}\sqrt{\frac{3}{2}}(D+F)$ & $\frac{1}{4}\left(D+F\right)$ & $2\sqrt{\frac{2}{3}}D$ & $\frac{1}{2}(D+2F)$ & $\frac{1}{2\sqrt{2}}(D+2F)$ & $\frac{1}{2}\sqrt{\frac{3}{2}}D$\tabularnewline
      $K_{3a}^{K}$ & $\frac{1}{2}\left(D+F\right)$ & $0$ & $D-F$ & $-\frac{D}{\sqrt{6}}$ & $\frac{D+F}{2}$ & $\frac{D+F}{2\sqrt{2}}$ & $-\frac{1}{2}\sqrt{\frac{3}{2}}(D-3F)$\tabularnewline
      $K_{3a}^{\eta}$ & $0$ & $\frac{1}{4}\sqrt{\frac{3}{2}}(D-3F)$ & $\frac{1}{4}(D-3F)$ & $0$ & $\frac{D}{2}$ & $\frac{D}{2\sqrt{2}}$ & $-\frac{1}{2}\sqrt{\frac{3}{2}}D$\tabularnewline
      \hline 
      $K_{3b}^{\pi}$ & $-D-F$ & $-\frac{1}{2}\sqrt{\frac{3}{2}}D$ & $\frac{1}{2}\left(2F-D\right)$ & $0$ & $-\frac{D-F}{4}$ & $-\frac{D-F}{4\sqrt{2}}$ & $\frac{3}{4}\sqrt{\frac{3}{2}}(D-F)$\tabularnewline
      $K_{3b}^{K}$ & $-\frac{1}{2}\left(D+F\right)$ & $\frac{1}{2}\sqrt{\frac{3}{2}}(D+3F)$ & $\frac{1}{2}(-D+F)$ & $-\sqrt{\frac{3}{2}}D$ & $-\left(D+F\right)$ & $-\frac{D+F}{\sqrt{2}}$ & $0$\tabularnewline
      $K_{3b}^{\eta}$ & $0$ & $\frac{1}{2}\sqrt{\frac{3}{2}}D$ & $-\frac{D}{2}$ & $0$ & $-\frac{(D+3F)}{4}$ & $-\frac{(D+3F)}{4\sqrt{2}}$ & $-\frac{1}{4}\sqrt{\frac{3}{2}}(D+3F)$\tabularnewline
      \hline 
      \hline 
    \end{tabular}
  \end{center}

  \begin{center}
    \begin{tabular}{cc}
      \hline 
      \hline 
      $K_{C3}:$ $n\to p$ & $8(h_{38}+h_{40})M_{K}^{2}-(4h_{38}-4h_{40}-8h_{44})M_{\pi}^{2}$\tabularnewline
      $K_{C3}:$ $\Lambda\to p$ & $-\sqrt{\frac{2}{3}}(8h_{38}+8h_{40}-4h_{41}-4h_{43}+8h_{44})M_{K}^{2}-\sqrt{\frac{2}{3}}(-4h_{38}-2h_{39}+4h_{40}-2h_{41})M_{\pi}^{2}$\tabularnewline
      $K_{C3}:$ $\Sigma^{-}\to n$ & $8(h_{41}+h_{43})M_{K}^{2}+4(h_{39}+h_{41})M_{\pi}^{2}$\tabularnewline
      $K_{C3}:$ $\Sigma^{-}\to\Lambda$ & $4\sqrt{\frac{2}{3}}(h_{40}+h_{41})M_{K}^{2}-\sqrt{\frac{2}{3}}(-2h_{38}-2h_{39}-2h_{40}-2h_{41}-4h_{43}-4h_{44})M_{\pi}^{2}$\tabularnewline
      $K_{C3}:$ $\Xi^{0}\to\Sigma^{+}$ & $8(h_{40}+h_{44})M_{K}^{2}+4(h_{38}+h_{40})M_{\pi}^{2}$\tabularnewline
      $K_{C3}:$ $\Xi^{-}\to\Sigma^{0}$ & $4\sqrt{2}(h_{40}+h_{44})M_{K}^{2}+2\sqrt{2}(h_{38}+h_{40})M_{\pi}^{2}$\tabularnewline
      $K_{C3}:$ $\Xi^{-}\to\Lambda$ & $-\sqrt{\frac{2}{3}}(8h_{39}-4h_{40}+8h_{41}+8h_{43}-4h_{44})M_{K}^{2}-\sqrt{\frac{2}{3}}(-2h_{38}-4h_{39}-2h_{40}+4h_{41})M_{\pi}^{2}$\tabularnewline
      \hline 
    \end{tabular}
  \end{center}
  
  \begin{center}
    \begin{tabular}{cccc}
      \hline 
      \hline 
      $K_{3}^{\phi}$ & $\phi=\pi$  & $\phi=K$ & $\phi=\eta$\tabularnewline
      \hline 
      $n\to p$ & $\frac{1}{4}\left(D+F\right)^{3}$ & $\frac{1}{3}\left(D^{3}-D^{2}F+3DF^{2}-3F^{3}\right)$ & $-\frac{1}{12}(D-3F)^{2}(D+F)$\tabularnewline
      $\Lambda\to p$ & $\frac{1}{2}\sqrt{\frac{3}{2}}D\left(-D^{2}+F^{2}\right)$ & $\frac{5D^{3}-15D^{2}F-9DF^{2}+27F^{3}}{6\sqrt{6}}$ & $\frac{D\left(D^{2}-9F^{2}\right)}{6\sqrt{6}}$\tabularnewline
      $\Sigma^{-}\to n$ & $\frac{1}{6}\left(D^{3}-2D^{2}F+3DF^{2}+6F^{3}\right)$ & $\frac{1}{6}\left(D^{3}+D^{2}F+3DF^{2}+3F^{3}\right)$ & $\frac{1}{6}D\left(D^{2}-4DF+3F^{2}\right)$\tabularnewline
      $\Sigma^{-}\to\Lambda$ & $-\frac{1}{3}\sqrt{\frac{2}{3}}D\left(D^{2}-6F^{2}\right)$ & $\frac{D\left(D^{2}-F^{2}\right)}{\sqrt{6}}$ & $\frac{1}{3}\sqrt{\frac{2}{3}}D^{3}$\tabularnewline
      $\Xi^{0}\to\Sigma^{+}$ & $\frac{D^{3}+2D^{2}F+3DF^{2}-6F^{3}}{6}$ & $\frac{D^{3}-D^{2}F+3DF^{2}-3F^{3}}{6}$ & $\frac{D\left(D^{2}+4DF+3F^{2}\right)}{6}$\tabularnewline
      $\Xi^{-}\to\Sigma^{0}$ & $\frac{D^{3}+2D^{2}F+3DF^{2}-6F^{3}}{6\sqrt{2}}$ & $\frac{D^{3}-D^{2}F+3DF^{2}-3F^{3}}{6\sqrt{2}}$ & $\frac{D\left(D^{2}+4DF+3F^{2}\right)}{6\sqrt{2}}$\tabularnewline
      $\Xi^{-}\to\Lambda$ & $\frac{1}{2}\sqrt{\frac{3}{2}}D\left(-D^{2}+F^{2}\right)$ & $\frac{5D^{3}+15D^{2}F-9DF^{2}-27F^{3}}{6\sqrt{6}}$ & $\frac{D\left(D^{2}-9F^{2}\right)}{6\sqrt{6}}$\tabularnewline
      \hline 
      \hline 
    \end{tabular}
  \end{center}
    
  \begin{center}
    \begin{tabular}{cccccccc}
      \hline 
      \hline 
      $K_{D}^{\phi}$ & $n\to p$ & $\Lambda\to p$ & $\Sigma^{-}\to n$ & $\Sigma^{-}\to\Lambda$ & $\Xi^{-0}\to\Sigma^{+}$ & $\Xi^{-}\to\Sigma^{0}$ & $\Xi^{-}\to\Lambda$\tabularnewline
      \hline 
      $K_{D3a}^{\pi}$ & $\frac{8}{3}\left(D+F\right)$ & $-\sqrt{\frac{3}{2}}(D+F)$ & $\frac{2}{3}(D+F)$ & $-\frac{2}{3}\sqrt{\frac{2}{3}}D$ & $\frac{2}{3}(D+2F)$ & $\frac{1}{3}\sqrt{2}(D+2F)$ & $-\sqrt{\frac{2}{3}}D$\tabularnewline
      $K_{D3a}^{K}$ & $\frac{1}{3}(3D+F)$ & $-2\sqrt{\frac{2}{3}}D$ & $\frac{4F}{3}$ & $\frac{1}{3}\sqrt{\frac{2}{3}}(5D+9F)$ & $\frac{7D+5F}{3}$ & $\frac{7D+5F}{3\sqrt{2}}$ & $\sqrt{\frac{3}{2}}(D-F)$\tabularnewline
      $K_{D3a}^{\eta}$ & $0$ & $0$ & $-\frac{1}{3}(D-3F)$ & $\sqrt{\frac{2}{3}}D$ & $\frac{2}{3}D$ & $\frac{\sqrt{2}}{3}D$ & $\sqrt{\frac{2}{3}}D$\tabularnewline
      \hline 
      $K_{D3b}^{\pi}$ & $\frac{8}{3}\left(D+F\right)$ & $-4\sqrt{\frac{2}{3}}D$ & $\frac{8F}{3}$ & $\sqrt{\frac{2}{3}}(D+2F)$ & $\frac{2}{3}(D-F)$ & $\frac{1}{3}\sqrt{2}(D-F)$ & $\sqrt{\frac{3}{2}}(D-F)$\tabularnewline
      $K_{D3b}^{K}$ & $\frac{1}{3}(3D+F)$ & $\frac{D-3F}{\sqrt{6}}$ & $\frac{1}{3}(D+F)$ & $\sqrt{\frac{2}{3}}(D+F)$ & $\frac{8}{3}(D+F)$ & $\frac{4}{3}\sqrt{2}(D+F)$ & $0$\tabularnewline
      $K_{D3b}^{\eta}$ & $0$ & $0$ & $0$ & $0$ & $\frac{D+3F}{3}$ & $\frac{D+3F}{3\sqrt{2}}$ & $0$\tabularnewline
      \hline 
      $K_{D3}^{\pi}$ & $\frac{20}{9}$ & $-2\sqrt{\frac{2}{3}}$ & $-\frac{4}{9}$ & $\frac{2\sqrt{\frac{2}{3}}}{3}$ & $\frac{4}{9}$ & $\frac{2\sqrt{2}}{9}$ & $\sqrt{\frac{2}{3}}$\tabularnewline
      $K_{D3}^{K}$ & $\frac{4}{9}$ & $-\sqrt{\frac{2}{3}}$ & $-\frac{2}{9}$ & $\frac{\sqrt{\frac{2}{3}}}{3}$ & $\frac{14}{9}$ & $\frac{7\sqrt{2}}{9}$ & $\sqrt{\frac{2}{3}}$\tabularnewline
      $K_{D3}^{\eta}$ & $0$ & $0$ & $0$ & $0$ & $\frac{2}{3}$ & $\frac{\sqrt{2}}{3}$ & $0$\tabularnewline
      \hline 
      \hline 
    \end{tabular}
  \end{center}
\end{table}

  \begin{table}[H]
    \caption{\label{tab:CoeffSelfEnergies} Coefficients of the self-energy graphs
      contributing to the octet baryon mass.}
    
    \begin{center}    
      \begin{tabular}{ccccc}
        \hline 
        \hline 
        $ $ & $N$ & $\Lambda$ & $\Sigma$ & $\Xi$\tabularnewline
        \hline 
        $K_{B}^{\phi}$ & $\frac{3}{4}\left(D+F\right)^{2}$ & $D^{2}$ & $\frac{1}{3}\left(D^{2}+6F^{2}\right)$ & $\frac{3}{4}\left(D-F\right)^{2}$\tabularnewline
        $K_{B}^{K}$ & $\frac{1}{6}\left(5D^{2}-6FD+9F^{2}\right)$ & $\frac{2}{3}\left(D^{2}+9F^{2}\right)$ & $\left(D^{2}+F^{2}\right)$ & $\frac{1}{6}\left(9F^{2}+6FD+5D^{2}\right)$\tabularnewline
        $K_{B}^{\eta}$ & $\frac{1}{12}\left(3F-D\right)^{2}$ & $\frac{1}{3}D^{2}$ & $\frac{1}{3}D^{2}$ & $\frac{1}{12}\left(3F+D\right)^{2}$\tabularnewline
        \hline 
        $K_{D}^{\pi}$ & $-4$ & $-3$ & $-\frac{2}{3}$ & $-1$\tabularnewline
        $K_{D}^{K}$ & $-1$ & $-2$ & $-\frac{10}{3}$ & $-3$\tabularnewline
        $K_{D}^{\eta}$ & $0$ & $0$ & $-1$ & $-1$\tabularnewline
        \hline 
        \hline 
      \end{tabular}
    \end{center}
  \end{table}
    
  \begin{table}[H]
    \caption{\label{tab:CoeffAV3} Coefficients for graphs $C_{1}$, $T_{3}$,
      $B_{3ab}$, $B_{3}$ $C_{3}$ and the decuplet graphs $D_{3ab}$ and
      $D_{3}$ contributing to the the axial-vector isovector baryon charges
      $g_{A,BB}^{\lambda=3}$. }
    
    \begin{center}
      \begin{tabular}{ccccccccc}
        \hline 
        \hline 
        & $p\to p$ & $\Sigma^{+}\to\Sigma^{+}$ & $\Xi^{0}\to\Xi^{0}$ &  &  & $p\to p$ & $\Sigma^{+}\to\Sigma^{+}$ & $\Xi^{0}\to\Xi^{0}$\tabularnewline
        \hline 
        $K_{C1}$ & $D+F$ & $2F$ & $-D+F$ &  &  &  &  & \tabularnewline
        \hline 
        $K_{T3}^{\pi}$ & $D+F$ & $2F$ & $-D+F$ &  & $K_{D3}^{\pi}$ & $\frac{20}{9}$ & $\frac{2}{9}$ & $-\frac{1}{9}$\tabularnewline
        $K_{T3}^{K}$ & $\frac{1}{2}\left(D+F\right)$ & $F$ & $\frac{1}{2}\left(-D+F\right)$ &  & $K_{D3}^{K}$ & $\frac{4}{9}$ & $\frac{22}{9}$ & $\frac{4}{9}$\tabularnewline
        $K_{T3}^{\eta}$ & $0$ & $0$ & $0$ &  & $K_{D3}^{\eta}$ & $0$ & $\frac{2}{3}$ & $\frac{1}{3}$\tabularnewline
        \hline 
        $K_{3a}^{\pi}$ & $D+F$ & $2F$ & $-D+F$ &  & $K_{D3a}^{\pi}$ & $\frac{8(D+F)}{3}$ & $\frac{2(D+F)}{3}$ & $\frac{D-F}{3}$\tabularnewline
        $K_{3a}^{K}$ & $\frac{1}{2}\left(D+F\right)$ & $F$ & $\frac{1}{2}\left(-D+F\right)$ &  & $K_{D3a}^{K}$ & $D+\frac{F}{3}$ & $2D-\frac{2F}{3}$ & $\frac{1}{3}(-D-5F)$\tabularnewline
        $K_{3a}^{\eta}$ & $0$ & $0$ & $0$ &  & $K_{D3a}^{\eta}$ & $0$ & $\frac{2D}{3}$ & $\frac{1}{3}(-D-3F)$\tabularnewline
        \hline 
        $K_{3b}^{\pi}$ & $-D-F$ & $-2F$ & $D-F$ &  & $K_{D3b}^{\pi}$ & $\frac{8(D+F)}{3}$ & $\frac{2(D+F)}{3}$ & $\frac{D-F}{3}$\tabularnewline
        $K_{3b}^{K}$ & $-\frac{1}{2}\left(D+F\right)$ & $-F$ & $\frac{1}{2}\left(D-F\right)$ &  & $K_{D3b}^{K}$ & $D+\frac{F}{3}$ & $2D-\frac{2F}{3}$ & $\frac{1}{3}(-D-5F)$\tabularnewline
        $K_{3b}^{\eta}$ & $0$ & $0$ & $0$ &  & $K_{D3b}^{\eta}$ & $0$ & $\frac{2D}{3}$ & $\frac{1}{3}(-D-3F)$\tabularnewline
        \hline 
        \hline 
      \end{tabular}
    \end{center}
    
    \begin{center}
      \begin{tabular}{cccc}
        \hline 
        \hline 
        $K_{3}^{\phi}$ & $\phi=\pi$  & $\phi=K$ & $\phi=\eta$\tabularnewline
        \hline 
        $p\to p$ & $\frac{1}{4}\left(D+F\right)^{3}$ & $\frac{1}{3}(D-F)(3F(-D+F)+D(D+3F))$ & $-\frac{1}{12}(D-3F)^{2}(D+F)$\tabularnewline
        $\Sigma^{+}\to\Sigma^{+}$ & $\frac{1}{3}\left(4D^{2}F-6F^{3}\right)$ & $F\left(D^{2}-F^{2}\right)$ & $-\frac{2D^{2}F}{3}$\tabularnewline
        $\Xi^{0}\to\Xi^{0}$ & $-\frac{1}{4}\left(D-F\right)^{3}$ & $-\frac{1}{3}(D+F)(D(D-3F)+3F(D+F))$ & $\frac{1}{12}(D-F)(D+3F)^{2}$\tabularnewline
        \hline 
        \hline 
      \end{tabular}
    \end{center}
    
    \begin{center}
      \begin{tabular}{cc}
        \hline 
        \hline 
        $K_{C3}:$ $p\to p$ & $8h_{44}M_{\pi}^{2}+4h_{38}\left(2M_{k}^{2}-M_{\pi}^{2}\right)+4h_{40}\left(2M_{k}^{2}+M_{\pi}^{2}\right)$\tabularnewline
        $K_{C3}:$ $\Sigma^{+}\to\Sigma^{+}$ & $4h_{38}M_{\pi}^{2}-4h_{39}M_{\pi}^{2}-8h_{43}M_{\pi}^{2}+8h_{44}M_{\pi}^{2}+4h_{40}\left(2M_{k}^{2}+M_{\pi}^{2}\right)-4h_{41}\left(2M_{k}^{2}+M_{\pi}^{2}\right)$\tabularnewline
        $K_{C3}:$ $\Xi^{0}\to\Xi^{0}$ & $-8h_{43}M_{\pi}^{2}-4h_{39}\left(2M_{k}^{2}-M_{\pi}^{2}\right)-4h_{41}\left(2M_{k}^{2}+M_{\pi}^{2}\right)$\tabularnewline
        \hline 
        \hline 
      \end{tabular}
    \end{center}
  \end{table}

\section{Heavy-baryon results\label{sec:Heavy-baryon-results}}

\begin{table}[H]
  \caption{Results for the heavy-baryon scheme. Notation is as in Tab. \ref{tab:SHD-contributions}
  }
  \begin{center}
    \begin{tabular}{cccccccccc}
      \hline 
      \hline 
      $D$ & $F$ & $F/D$ & $h_{38}$ [GeV$^{-2}$] & $h_{39}$ [GeV$^{-2}$] & $h_{41}$ [GeV$^{-2}$] & $h_{41}$ [GeV$^{-2}$] & $h_{43}$ [GeV$^{-2}$] & $h_{44}$ [GeV$^{-2}$] & $\chi_{\mbox{red}}^{2}$\tabularnewline
      $0.626(57)$ & $0.465(48)$ & $0.71$ & $0.053(19)$ & $-0.007(30)$ & $-0.175(77)$ & $0.002(41)$ & $0.044(33)$ & $-0.159(21)$ & $\frac{6.8}{11}=0.62$\tabularnewline
      \hline 
      \hline 
    \end{tabular}
  \end{center}
  
  \begin{center}
    \begin{tabular}{cccccccccccc}
      \hline 
      \hline 
      & $g_{1}$ & $n\to p$ & $\Lambda\to p$ & $\Sigma^{-}\to n$ & $\Sigma^{-}\to\Lambda$ & $\Xi^{0}\to\Sigma^{+}$ & $\Xi^{-}\to\Lambda$ & $\Sigma^{_-}\to\Sigma^{_{0}}$ & $\Xi^{-}\to\Xi^{0}$ & $g_{A,3}^{\Sigma^+}$ & $g_{A,3}^{\Xi^0}$\tabularnewline
      \hline 
      Exp &  & $1.270\left(3\right)$ & $-0.879\left(18\right)$ & $0.340\left(17\right)$ & $0.588\left(16\right)$ & $1.210\left(50\right)$ & $0.306\left(61\right)$ &  &  &  & \tabularnewline
      \hline 
      HB & LO & $1.09$ & $-0.82$ & $0.16$ & $0.51$ & $1.09$ & $0.31$ & $0.66$ & $0.16$ & $0.93$ & $-0.16$\tabularnewline
      & $C_{3}$ & $-0.28$ & $0.50$ & $0.08$ & $-0.15$ & $-0.67$ & $-0.33$ & $-0.27$ & $-0.01$ & $-0.38$ & $0.01$\tabularnewline
      & $T_{3}$ & $0.26$ & $-0.37$ & $0.07$ & $0.12$ & $0.49$ & $0.14$ & $0.15$ & $0.04$ & $0.22$ & $-0.04$\tabularnewline
      & $B^{loops}$ & $0.21$ & $-0.06$ & $0.02$ & $0.12$ & $0.30$ & $0.10$ & $0.17$ & $0.02$ & $0.23$ & $-0.02$\tabularnewline
      & full & $1.270(3)$ & $-0.886(18)$ & $0.330(16)$ & $0.599(22)$ & $1.210(39)$ & $0.228(42)$ & $0.707(52)$ & $0.212(36)$ & $1.000(73)$ & $-0.212(36)$\tabularnewline
      & $|p^{3}/p^{1}|$ & $0.16$ & $0.07$ & $1.04$ & $0.17$ & $0.11$ & $0.27$ & $0.08$ & $0.31$ & $0.08$ & $0.31$\tabularnewline
      \hline 
      \hline 
    \end{tabular}
  \end{center}
\end{table}

\end{appendix}

\end{document}